\documentclass[fleqn,usenatbib]{mnras}

\usepackage{newtxtext,newtxmath}

\usepackage[T1]{fontenc}
\usepackage{ae,aecompl}

\usepackage[dvipdfmx]{graphicx}	
\usepackage{amsmath}	
\usepackage{amssymb}	
\usepackage{threeparttable}
\usepackage{url}

\title[Galaxy Proto-Cluster Cores at $z\sim2$]{A Systematic Search for Galaxy Proto-Cluster Cores at $z\sim 2$}

\author[M. Ando, K. Shimasaku and R. Momose]{
Makoto Ando,$^{1}$\thanks{E-mail: mando@astron.s.u-tokyo.ac.jp}
Kazuhiro Shimasaku$^{1,2}$ and Rieko Momose$^{1}$
\\
$^{1}$Department of Astronomy, Graduate School of Science, The University of Tokyo, 7-3-1 Hongo, Bunkyo-ku, Tokyo 113-0033, Japan\\
$^{2}$Research Center for the Early Universe, The University of Tokyo, 7-3-1 Hongo, Bunkyo-ku, Tokyo 113-0033, Japan
}

\date{Accepted XXX. Received YYY; in original form ZZZ}

\pubyear{2020}

\begin{document}
\label{firstpage}
\pagerange{\pageref{firstpage}--\pageref{lastpage}}
\maketitle

\begin{abstract}
A proto-cluster core is the most massive dark matter halo (DMH) in a given proto-cluster. To reveal the galaxy formation in core regions, we search for proto-cluster cores at $z\sim 2$ in $\sim 1.5\, \mathrm{deg}^{2}$ of the COSMOS field. Using pairs of massive galaxies ($\log(M_{*}/M_{\sun})\geq11$) as tracers of cores, we find 75 candidate cores, among which 54\% are estimated to be real. A clustering analysis finds that these cores have an average DMH mass of $2.6_{-0.8}^{+0.9}\times 10^{13}\, M_{\sun}$, or $4.0_{-1.5}^{+1.8}\, \times 10^{13} M_{\sun}$ after contamination correction. The extended Press-Schechter model shows that their descendant mass at $z=0$ is consistent with Fornax-like or Virgo-like clusters. Moreover, using the IllustrisTNG simulation, we confirm that pairs of massive galaxies are good tracers of DMHs massive enough to be regarded as proto-cluster cores. We then derive the stellar mass function (SMF) and the quiescent fraction for member galaxies of the 75 candidate cores. We find that the core galaxies have a more top-heavy SMF than field galaxies at the same redshift, showing an excess at $\log(M_{*}/M_{\sun})\gtrsim 10.5$. The quiescent fraction, $0.17_{-0.04}^{+0.04}$ in the mass range $9.0\leq \log(M_{*}/M_{\sun})\leq 11.0$, is about three times higher than that of field counterparts, giving an environmental quenching efficiency of $0.13_{-0.04}^{+0.04}$. These results suggest that stellar mass assembly and quenching are accelerated as early as $z\sim 2$ in proto-cluster cores.
\end{abstract}




\section{Introduction}
In the $\rm \Lambda$CDM framework, the initial density perturbations grow by gravity and form dark matter haloes (DMHs), and galaxies are formed in DMHs through gas cooling. DMHs, and hence galaxies, become more massive and larger through the accretion of matter and mergers with other DMHs. The most massive and largest DMHs in today's universe are galaxy clusters. The DMH mass of galaxy clusters is typically $\gtrsim 10^{14}\, M_{\sun}$ (e.g. \citealp{Kravtsov_Borgani_2012,Overzier2016}) and a mature cluster hosts hundreds to thousands of galaxies.

The properties of cluster galaxies are largely different from those of field galaxies. For example, at $z<1$, cluster galaxies are dominated by quiescent and/or elliptical galaxies with old stellar populations while most field galaxies are star-forming galaxies like spirals (e.g. \citealp{Dressler1980,Goto2003,Bower1998}).  Part or all of these differences are thought to be caused by some environmental effects: ram-pressure stripping \citep{Gunn1972}, galaxy interaction, galaxy harassment \citep{Moore1998}, etc. When and how these differences were established is key to understanding the role of environments on galaxy formation. For this purpose, galaxies in clusters in early evolutionary stages should be investigated.

Progenitors of local clusters at $z\gtrsim2$ are called proto-clusters. They are defined as a whole structure that will collapse into a cluster by $z=0$ (e.g. \citealp{Overzier2016}). A Proto-cluster typically extends to more than 20 comoving Mpc at $z\sim 2$ \citep{Chiang2013,Muldrew2015} and an even larger area at higher redshift, being split into a number of DMHs and unbound regions. Among those substructures, we define the ``core" of the proto-cluster as the most massive DMH\footnote{Massive systems with $M_\mathrm{DMH}>10^{14}\,M_{\sun}$ at $z\gtrsim 2$ are sometimes called high redshift clusters. However, they will also grow through the accretion of matter from the surrounding regions until $z=0$. In this sense, they are also regarded as massive proto-cluster cores.}. The mass of cores has a large scatter ($\sim 1\,\mathrm{dex}$) at $z\sim 2$, even if the descendant mass at $z=0$ is fixed \citep{Muldrew2015}.

The relationship between the properties of galaxies and their location in proto-clusters is important to understand cluster galaxy formation. \citet{Muldrew2018} have studied galaxy evolution in proto-clusters by applying a semi-analytic galaxy evolution model to N-body simulations. They have found that galaxies in core regions have different properties from those in fields and the rest of the proto-cluster regions: a more top-heavy stellar mass function (SMF) and a higher fraction of quiescent galaxies especially for low-mass galaxies. A similar trend of the SMF has been reported in \citet{Muldrew2015,Lovell2018}.

Recently, several proto-cluster cores have been found and they have a variety of star formation activity. \citet{Shimakawa2018} have found that $\mathrm{H \alpha}$ emitters in the densest regions of a proto-cluster at $z\sim 2.5$, which are regarded as cores, are more massive and more actively star-forming than those in the remaining regions of the same proto-cluster. Some cores are dominated by (dusty) star-forming galaxies unlike local mature clusters \citep{Wang2016, Oteo2018, Miller2018}, while massive cores with red sequence galaxies, which are similar to local clusters, have also been found \citep{Newman2014,Cooke2016,Lee-Brown2017,Willis2020}. Such variations may reflect different evolutionary stages of cores. Most of the reported cores are biased to possible progenitors of the most massive, Coma-like clusters ($M_\mathrm{DMH}>10^{15}\, M_{\sun}$ by $z=0$). Therefore, to reveal the whole aspect of galaxy evolution in proto-cluster cores, we need a large sample of cores including less massive ones.

Systematic proto-cluster searches have been done by various techniques. One of such methods, the fixed aperture method, searches for an overdensity of high redshift galaxies (i.e. Lyman break galaxies (LBGs), line emitters, photometric redshift galaxies, etc.) over a given aperture (e.g. \citealp{Chiang2014,Toshikawa2018}). This method can successfully identify the whole region of a proto-cluster (e.g. \citealp{Chiang2015,Diener2015}). However, because this method uses a ten times larger aperture than the size of cores, it is difficult to isolate cores. Moreover, because LBGs and line emitters are typically star-forming galaxies, overdensities of such populations provide a biased view of proto-cluster galaxies.

Another method is to use biased tracers. Some galaxy populations like high redshift radio galaxies and quasars are frequently located at dense environments \citep{Hatch2011,Hatch2014}. Therefore, one can use such objects as beacons of proto-clusters (\citealp{Venemans2007,Wylezalek2013,Wylezalek2014,Cooke2014}). However, it is unclear whether these objects can trace proto-clusters completely \citep{Lovell2018, Uchiyama2018}. Because the lifetime of quasars, $10^{6}$ to $10^{8}$ years \citep{Martini2004}, is relatively short, they may miss some fraction of proto-clusters. Furthermore, the feedback of active galaxies suppresses the formation of surrounding galaxies \citep{Uchiyama2019}, possibly resulting in a biased picture of galaxy formation in proto-clusters.

In this study, we propose a new method to find proto-cluster cores at $z\sim 2$, the epoch when massive cores appear \citep{Chiang2017}, and use it in the Cosmic Evolution Survey (COSMOS; \citealp{Scoville2007}) field. The extended Press-Schechter model\footnote{To calculate the extended Press-Schechter model, we use a FORTRAN code written by Takashi Hamana. The code is found at \url{http://th.nao.ac.jp/MEMBER/hamanatk/OPENPRO/index.html}} predicts that a DMH whose mass is $\gtrsim 2-3 \times 10^{13}\, M_{\sun}$ at $z\sim 2$ typically evolves into the cluster mass regime, $\gtrsim10^{14}\, M_{\sun}$, by $z=0$. Therefore, we regard DMHs with $\gtrsim 2-3 \times 10^{13}\, M_{\sun}$ at $z\sim 2$ as proto-cluster cores and search for such massive systems.

The stellar to halo mass relation says that galaxies with larger stellar masses are hosted by more massive DMHs. According to abundance matching technique, the typical stellar mass of central galaxies hosted by DMHs with $M_{\mathrm{DMH}}\gtrsim 10^{13}\, M_{\sun}$ is $M_{*}\gtrsim 10^{11}\, M_{\sun}$ (e.g. \citealp{Behroozi2013}). However, DMHs which host central galaxies with $M_{*}\gtrsim 10^{11}\, M_{\sun}$ cover a wide range of DMH mass ($10^{12}-10^{14}\, M_{\sun}$). This means that using single massive galaxies cannot isolate DMHs as massive as proto-cluster cores. 

A multiple system of massive galaxies is a possible tracer of a proto-cluster core. \citet{Bethermin2014} have studied the clustering of BzK-selected galaxies at $1.5<z<2.5$, finding that close pairs (separations are below $20\arcsec$) of massive ($M_{*}>10^{11}\, M_{\sun}$) quiescent galaxies as well as massive main-sequence galaxies with strong star formation ($>200\, M_{\sun}/\mathrm{yr}$) are possible progenitors of clusters. The host DMH masses of the former at $z\sim 2$ is $5.5_{-4.5}^{+5.1}\times 10^{13}\, M_{\sun}$, which is massive enough to be regarded as cores. Using a galaxy sample with spectroscopic redshifts, \citet{Diener2013} have explored candidate galaxy groups within $500\, \mathrm{kpc}$ in projected distance and $700\, \mathrm{km/s}$ in velocity difference at $1.8<z<3.0$. In comparison with mock galaxy catalogues, they have found that the candidate groups contain one thirds of the progenitors of present-day clusters, although they are mainly the progenitors of less massive systems ($10^{13}-10^{14}\, M_{\sun}$). Moreover, there is a significant overdensity not only of the spectroscopic redshift sample but also of a photometric redshift sample with $M_{*}\geq 10^{10}\, M_{\sun}$ around the candidate groups.

These results lead to an assumption that pairs of massive galaxies are hosted by more massive DMHs than isolated massive galaxies. Thus, we use a pair of massive galaxies as a tracer of proto-cluster cores. We define the term ``pair" as a multiple system of massive galaxies whose extent is consistent with the size of a proto-cluster core. We refer to not only associations of two massive galaxies but also those of more than two as ``pairs". Since most of the ``pairs" identified in this paper consist of just two galaxies, we adopt this naming convention. To avoid possible selection bias, we use both star-forming galaxies and quiescent galaxies to find out pairs.

The structure of this paper is as follows. In Section 2, we describe the data and galaxy samples used in this study. In Section 3, we introduce the method to find proto-cluster cores and show results. We also compare the results with the IllustrisTNG simulation to evaluate the effectiveness of our method. In Section 4, we examine properties of member galaxies in the core candidates focusing on the stellar mass function and the fraction of quiescent galaxies. Section 5 is  devoted to a summary and conclusions.
 
Throughout this paper, we assume a flat $\mathrm{\Lambda}$CDM cosmology with $(\Omega_\mathrm{m},\, \Omega_\mathrm{\Lambda},\, h,\, \sigma_\mathrm{8},\, n_{0})=(0.3,\, 0.7,\, 0.81, \,0.7,\, 0.9)$. We use the notations $\rm cMpc$ and $\rm pMpc$ to indicate comoving and physical scales, respectively.
We assume a \citet{Chabrier2003} initial mass function.

\section{Data and sample selection}
\subsection{The COSMOS2015 catalogue}
We use data from the COSMOS2015 galaxy catalogue (\citealp{Laigle2016}; COSMOS2015 hereafter). COSMOS2015 contains deep and multi-wavelength photometry, from near-ultraviolet (NUV) to far-infrared, on the COSMOS field. 

In this paper, we only use objects in the central $\sim 1.5\, \mathrm{deg}^{2}$ region covered by the UltraVISTA-DR2. We also limit our sample to galaxies with $m(K_\mathrm{s})\leq 24.0$. This magnitude cut is motivated so that the detection completeness is homogeneous over the UltraVISTA field.

From the catalogue, we extract the following quantities: photometric redshift (photo-\textit{z}), stellar mass and galaxy classification flag, an indicator of star formation activity (star-forming or quiescent). In the catalogue, the \texttt{LEPHARE} code \citep{Arnouts2002,Ilbert2006} has been used to compute photo-\textit{z}'s and perform spectral energy distribution (SED) fitting, and the \textit{NUV-r} vs \textit{r-J} colour-colour plane has been used to classify galaxies \citep{Williams2009}: quiescent galaxies are defined as those with $M_\mathrm{NUV}-M_\mathrm{r}>3(M_\mathrm{r}-M_\mathrm{J})+1$ and $M_\mathrm{NUV}-M_\mathrm{r}>3.1$. 

At $1.5\leq z\leq 3.0$, this parent sample contains 60080 (167844) galaxies in total after (before) the magnitude cut. Among them, 57353 galaxies are classified as star-forming galaxies while the remaining 2727 galaxies are quiescent galaxies.

\subsection{Sample selection}
\label{sec:sample_selection}
To identify massive DMHs, we use 1742 galaxies with $1.5\leq z\leq 3.0$ and $\log(M_{*}/M_{\sun})\geq 11$. We refer to the galaxies in this sample as ``massive galaxies (MGs)" hereafter. This sample accounts for 3\% of the parent sample at $1.5\leq z\leq 3.0$.

For the cross-correlation analysis described in Section~\ref{sec:clustering}, a relatively large galaxy sample is needed. From the COSMOS2015 catalogue, we select galaxies with $1.5\leq z\leq 3.0$ and $10.2<\log(M_{*}/M_{\sun})<11$, whose total number is 16149. We refer to the galaxies in this sample as ``tracer galaxies" hereafter. This sample accounts for 27\% of the parent sample at $1.5\leq z\leq 3.0$.

To examine properties of member galaxies of proto-cluster cores, we use all 86374 galaxies at $1.25\leq z\leq 3.25$. We refer to the galaxies in this sample as ``general galaxies" hereafter. These samples are summarised in Table~\ref{tab:sample}.

\begin{table*}
\centering
    \caption{Galaxy samples used in this paper.}
    \label{tab:sample}
	\begin{tabular}{lccccc}
		\hline
		sample & redshift & stellar mass cut & total & star-forming & quiescent \\
		        &          & $[M_{\sun}]$ & [\#] & [\#] & [\#] \\
		\hline
		parent sample & $1.5\leq z\leq 3.0$ & - & $60080$ & $57353$ & $2727$  \\  [2pt]
		massive galaxies (MGs) & $1.5\leq z\leq 3.0$ & $M_{*}\geq 10^{11}$ & $1742$ & $1207$ & $535$  \\  [2pt]
		tracer galaxies & $1.5\leq z\leq 3.0$ & $10^{10.2}< M_{*}<10^{11}$ & $16149$ & $14329$ & $1820$  \\  [2pt]
		\hline
		general galaxies & $1.25\leq z\leq 3.25$ & - & $ 86374$ & $82012$ & $4362$  \\
		\hline
	\end{tabular}
	\begin{tablenotes}[normal]
	 \item \textit{Note.} A magnitude cut of $m(K_\mathrm{s})\leq 24.0$ is applied to all samples.
    \end{tablenotes}
\end{table*}

\section{Construction of a proto-cluster core sample}
In this section, we describe the method to identify proto-cluster core candidates and how to estimate their DMH mass.

\subsection{Candidates for proto-cluster cores}
\subsubsection{Pair finder}
\label{sec:pair_finder}
We use pairs of MGs to search for proto-cluster cores. In this study, a ``pair of MGs" refers to a multiple system of MGs, whose size is consistent with that of proto-cluster cores, $\sim 0.3\, \mathrm{pMpc}$. To identify such systems, we apply the following procedure to the MGs:

\begin{enumerate}
  \item We pick up one galaxy and count neighbour galaxies within $\Delta \theta \leq 30\arcsec$ and $\Delta z \leq 0.12$ from that galaxy.\\
  \item  If the number of neighbours is more than one, we regard all of them as member galaxies of a ``pair".\\
  \item The three dimensional position of the pair is defined as the average position of the member galaxies of the pair.
\end{enumerate}
We set $30\arcsec$ as the maximum separation of member galaxies. This value is slightly smaller than the size of a core with $M_\mathrm{DMH}\sim 2\times 10^{13}\, M_{\sun}$, $\sim 36\arcsec \sim 0.3\, \mathrm{pMpc}$ in radius, reducing the probability of chance projection. We also set $0.12$ as the maximum redshift difference among members, considering the uncertainty in photo-\textit{z} estimates in the COSMOS2015 catalogue. Since $\Delta z =0.12$ corresponds to about $170\, \mathrm{cMpc}$ at $z=2$, which is much larger than the size of a core, detected pairs may be contaminated by false pairs due to chance projection. We discuss this in Section~\ref{sec:true_pair_fraction}.

\subsubsection{Detected core candidates}
\label{sec:pair}
\begin{figure}
	\includegraphics[width=\columnwidth]{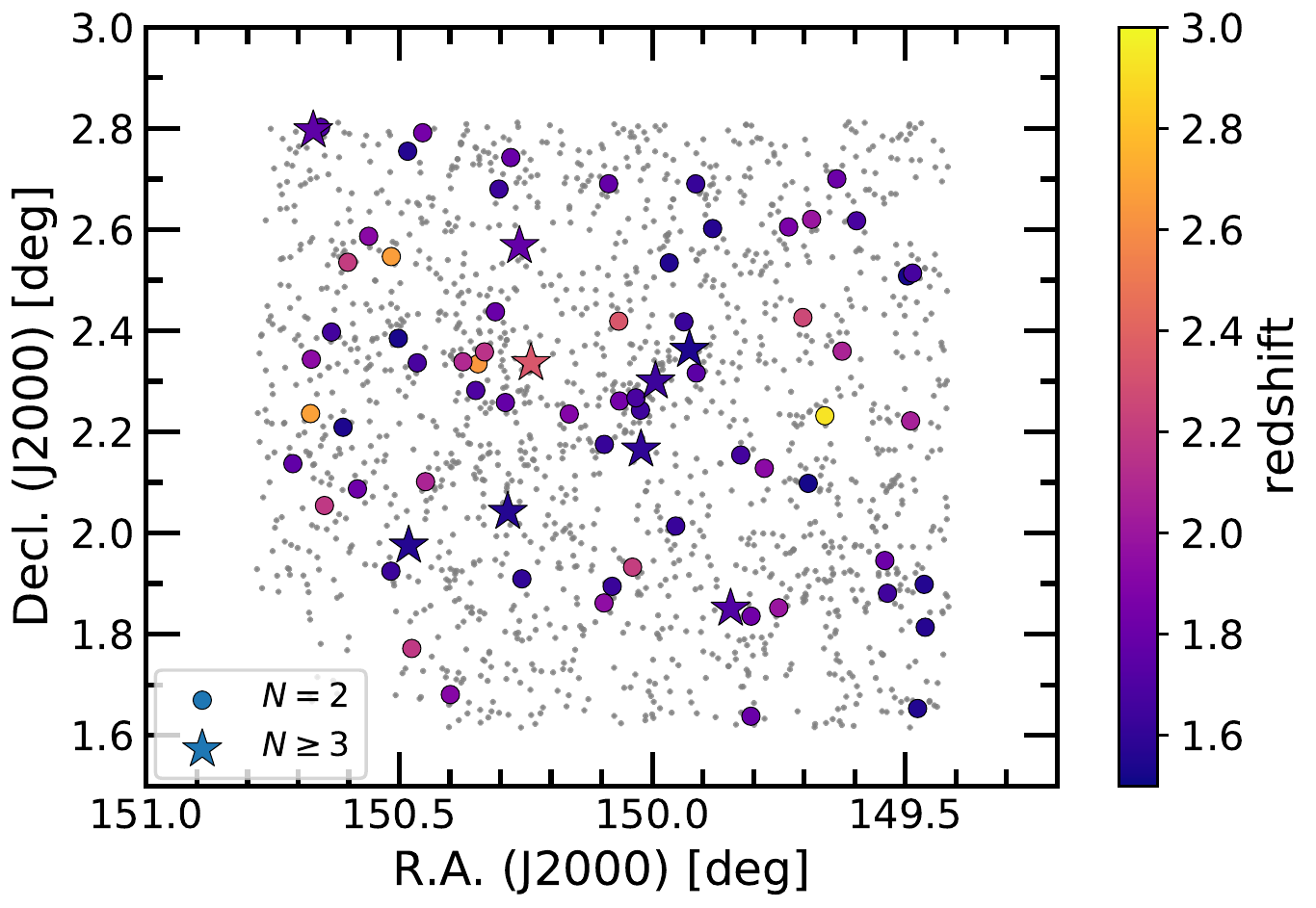}
    \caption{Sky distribution of the pairs of MGs colour-coded by redshift according to the colour bar. Pairs containing only two MGs are plotted as circles (66 pairs), while those containing three or four are plotted as star symbols (9 pairs). Grey dots are MGs (1742 in total). (A colour version of this figure is available in the online journal.)}
    \label{fig:pair_skydist}
\end{figure}

\begin{figure}
	\includegraphics[width=\columnwidth]{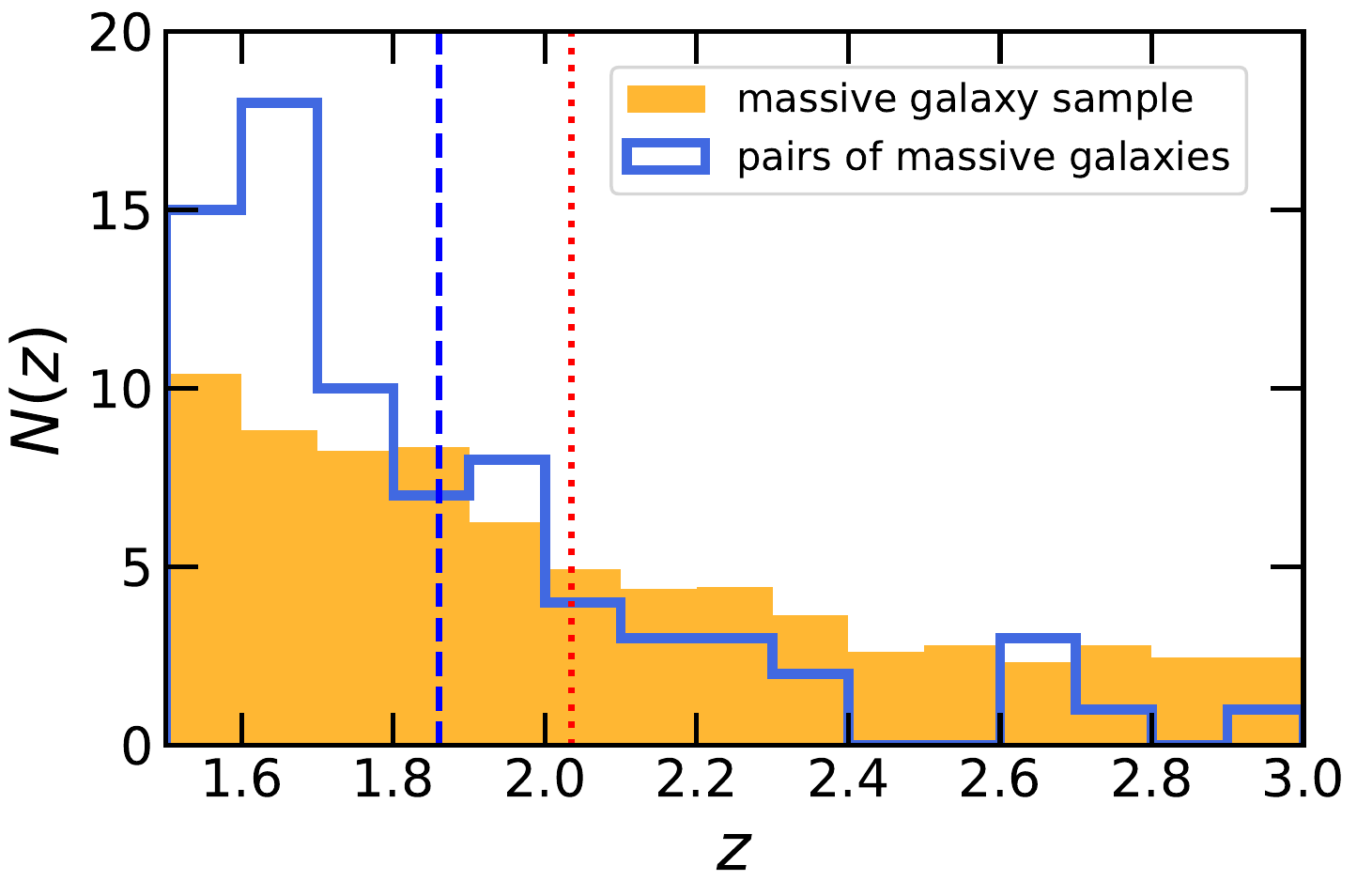}
    \caption{The redshift distribution of MGs (orange) and pairs (blue). The histogram of MGs is normalised so that the total number matches that of pairs. The average redshifts of MGs and pairs are shown by dotted and dashed lines, respectively. (A colour version of this figure is available in the online journal.)}
    \label{fig:pair_Nz}
\end{figure}

Applying our pair finder to the 1742 MGs, we identify 75 pairs as proto-cluster core candidates. Their sky position is shown in Fig.~\ref{fig:pair_skydist}. While the majority (66 pairs) have only two MGs, 9 pairs have three or four members, plotted as star symbols. The redshift distribution of the 75 pairs is shown in Fig.~\ref{fig:pair_Nz} with that of the MG sample. The average redshift of the pairs, 1.85, is lower than that of the MG sample, 2.03. This difference may reflect the fact that there are more massive virialized systems at lower redshifts. 

We note that our core candidates contain a very massive ($M_\mathrm{DMH}\sim 10^{14}\, M_{\sun}$) core at $z\sim 2.5$ which has been spectroscopically confirmed in \citet{Wang2016}.

\subsection{Clustering analysis}
\label{sec:clustering}
We use clustering analysis to estimate the average DMH mass of the core candidates obtained in Section~\ref{sec:pair}. Since we have only 75 candidates, we apply a cross-correlation technique.

\subsubsection{The auto-correlation function of tracer galaxies}
We first calculate the two-point angular auto-correlation function (ACF) of the tracer galaxy sample. We use an estimator of the ACF proposed by \citet{Landy1993}:
\begin{equation}
\label{ACF_estimator}
    \omega_\mathrm{ACF}(\theta)=\frac{DD(\theta)-2DR(\theta)+RR(\theta)}{RR(\theta)},
\end{equation}
where $DD(\theta),\ DR(\theta)$, and $RR(\theta)$ are the normalised number counts of galaxy-galaxy, galaxy-random, and random-random pairs whose separations are $\theta$, respectively. We use $2\times 10^{5}$ random points uniformly distributed over the entire area where the data exist. We assume that the errors in the ACF come from the Poisson error in the $DD(\theta)$ term,
\begin{equation}
    \varepsilon_\mathrm{ACF}=\frac{1+\omega_\mathrm{ACF}(\theta)}{\sqrt{DD_{0}(\theta)}},
\end{equation}
where $DD_{0}(\theta)$ is the row number count of galaxy-galaxy pairs. We assume that the ACF can be described by a power-law:
\begin{equation}
\label{omega_model}
    \omega_\mathrm{model}(\theta)=A_{\omega}\theta^{-\beta},
\end{equation}
where $A_{\omega}=\omega(1\arcsec)$ is the amplitude of the ACF. We fix $\beta$ to the fiducial value 0.8 (e.g. \citealp{Peebles1975,Ouchi2003}).

When we apply the estimator in Equation~\eqref{ACF_estimator} to observational data of a finite survey area, the ACF is negatively biased due to the integral constraint (IC; \citealp{Groth1977}):
\begin{equation}
    \omega_\mathrm{obs}(\theta) = \omega_\mathrm{true}(\theta) - \mathrm{IC},
\end{equation}
where $\omega_\mathrm{obs}$ is the ACF derived from the observational data and $\omega_\mathrm{true}$ is the true ACF. Following \citet{Roche1999}, we calculate this term using random points:
\begin{equation}
    \mathrm{IC} = \frac{\sum_{\theta} RR(\theta)\cdot \omega_\mathrm{model}(\theta)}{\sum_{\theta} RR(\theta)}=\frac{\sum_{\theta} RR(\theta)\cdot A_{\omega}\theta^{-\beta}}{\sum_{\theta} RR(\theta)}.
\end{equation}
We derive $\mathrm{IC} = 0.0027A_{\omega}$ in the COSMOS field.
We fit $\omega(\theta)$ over $40\arcsec-2000\arcsec$ with correction of the IC.

We then calculate the spatial two-point correlation function $\xi(r)$:
\begin{equation}
    \xi(r) = \left(\frac{r}{r_{0}}\right)^{-\gamma},
\end{equation}
where $r_{0}$ is the correlation length and $\gamma$ is slope of the power-law. The spatial correlation function $\xi(r)$ is linked to the angular correlation function $\omega(\theta)$ via the Limber transform \citep{Peebles1980,Efstathiou1991}:
\begin{align}
    \beta &= \gamma - 1,\\
\label{a_omega_limber}
    A_{\omega} &= r_{0}^{\gamma} B \left(\frac{1}{2}, \frac{\gamma-1}{2} \right) \frac{\int_{0}^{\infty}{dz N(z)^{2}F(z)D_{\theta}(z)^{1-\gamma}g(z)}}{\left[\int_{0}^{\infty}{dz N(z)}\right]^{2}},\\
    g(z) &= \frac{H_{0}}{c}(1+z)^{2}\left\{1+\Omega_\mathrm{m}z+\Omega_\mathrm{\Lambda}[(1+z)^{-2}-1]\right\}^{1/2},
\end{align}
where $B$ is the beta function, $N(z)$ is the redshift distribution of galaxies used to derive the ACF and $D_{\theta}(z)$ is the angular diameter distance. $F(z)$ describes the redshift evolution of $\xi(r)$, which is modelled as $F(z)=[(1+z)/(1+\Bar{z})]^{-(3+\Bar{\epsilon})}$ with $\Bar{\epsilon}=-1.2$ \citep{Roche1999}, where $\Bar{z}$ is the average redshift of the sample.

Then we define the linear bias parameter of galaxies $b_\mathrm{g}$, which represents the relative strength of galaxy clustering compared to dark matter at a large scale ($8\, \mathrm{cMpc}/{\it h_\mathrm{100}}$):
\begin{equation}
\label{bias}
    b_\mathrm{g} = \sqrt{\frac{\xi_\mathrm{g}\left(r=8\, \mathrm{cMpc}/ {\it h_\mathrm{100}}\right)}{\xi_\mathrm{DM}\left(r=8\, \mathrm{cMpc}/{\it h_\mathrm{100}}\right)}},
\end{equation}
where $\xi_\mathrm{DM}(r)$ is the spatial correlation function of dark matter. We assume \citet{Eisenstein1999} model as the power spectrum of matter. To calculate $\xi_\mathrm{DM}(r)$, we use a python toolkit for cosmological calculations called \texttt{COLOSSUS} \citep{Diemer2018}. In this way, the bias parameter of the tracer galaxies is derived from Equation~\eqref{bias}. We assume that the bias parameter of galaxies approximates that of the underlying DMHs on large scales.

\subsubsection{The cross-correlation function between cores and tracers}
Cross-correlation technique is often applied when the sample size is small. We calculate the two-point angular cross-correlation function (CCF) between the core candidates and the tracer galaxies using the following estimator:
\begin{equation}
\label{CCF_estimator}
    \omega_\mathrm{CCF}(\theta)=\frac{D_\mathrm{s}D_\mathrm{t}(\theta)-D_\mathrm{s}R(\theta)-D_\mathrm{t}R(\theta)+RR(\theta)}{RR(\theta)},
\end{equation}
where $D_\mathrm{s}D_\mathrm{t}(\theta)$, $D_\mathrm{s}R(\theta)$ and $D_\mathrm{t}R(\theta)$ are the normalised number counts of pair-tracer, pair-random, and tracer-random pairs whose separations are $\theta$, respectively. Since the sample sizes of tracers and random points are much larger than that of pairs, we assume that the errors in the CCF come from the Poisson error in the $D_\mathrm{s}D_\mathrm{t}(\theta)$ term:
\begin{equation}
    \varepsilon_\mathrm{CCF}=\frac{1+\omega_\mathrm{CCF}(\theta)}{\sqrt{D_\mathrm{s}D_{\mathrm{t}_{0}}(\theta)}},
\end{equation}
where $D_\mathrm{s}D_\mathrm{{t}_{0}}(\theta)$ is the row number count of pair-tracers. We fit the CCF using Equation~\eqref{omega_model} and derive its amplitude. Then, we calculate the correlation length of the spatial CCF in almost the same way as for the ACF. Instead of Equation~\eqref{a_omega_limber}, we use the following equation \citep{Croom1999}:
\begin{equation}
     A_{\omega} = r_{0}^{\gamma} B \left(\frac{1}{2}, \frac{\gamma-1}{2} \right) \frac{\int_{0}^{\infty}{dz N_\mathrm{s}(z)N_\mathrm{t}(z)F(z)D_{\theta}(z)^{1-\gamma}g(z)}}{\left[\int_{0}^{\infty}{dz N_\mathrm{s}(z)}\right] \cdot \left[\int_{0}^{\infty}{dz N_\mathrm{t}(z)}\right]},
\end{equation}
where $N_\mathrm{s}$ and $N_\mathrm{t}$ are the redshift distributions of pairs and tracer galaxies, respectively. For the term $F(z)$, we use the average redshift of pairs. After that, we derive the bias parameter of the cross-correlation from Equation~\eqref{bias}.

With the bias parameters of tracer galaxies ($b_\mathrm{t}$) and the cross-correlation ($b_\mathrm{st}$), we estimate that of core candidates by:
\begin{equation}
    b_\mathrm{s} = \frac{b_\mathrm{st}^{2}}{b_\mathrm{t}}.
\end{equation}

We use $b_\mathrm{s}$ to calculate the average mass of the core-hosting DMHs with the relation between the bias parameter $b$ and the ``peak height" in the linear density field, $\nu$, presented in \citet{Tinker2010}. Here, the peak height $\nu$ is defined as:
\begin{equation}
    \nu = \frac{\delta_\mathrm{c}}{\sigma(M)},
\end{equation}
where $\delta_\mathrm{c}= 1.686$ is the critical density for spherical collapse, and $\sigma(M)$ is the linear matter standard deviation on the Lagrangian scale of the halo. For this calculation, we use the python toolkit \texttt{COLOSSUS}.

\subsubsection{DMH mass of the core candidates}
\label{sec:DMH_mass}
\begin{figure}
	\includegraphics[width=\columnwidth]{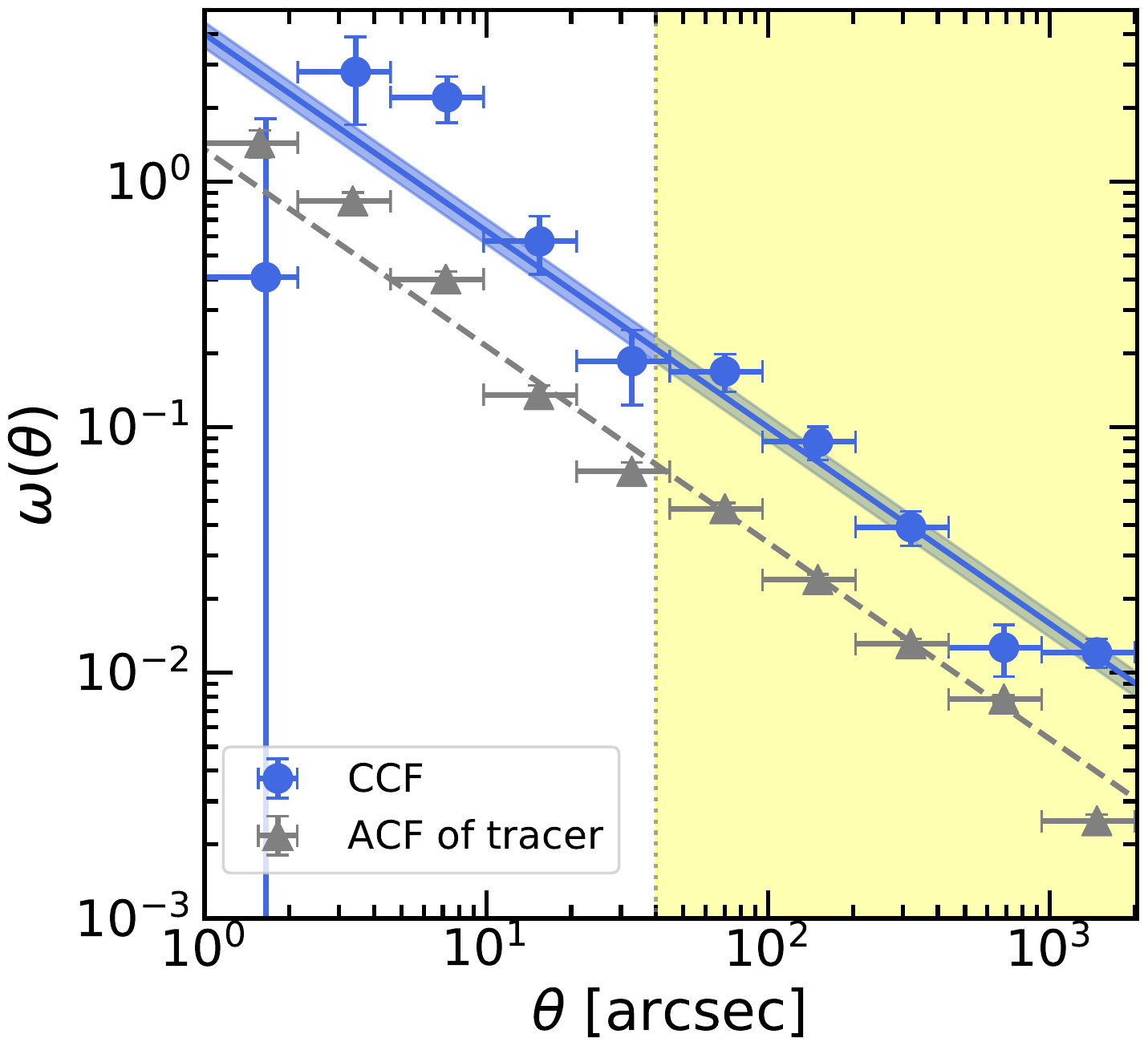}
    \caption{The ACF of tracer galaxies (grey) and the CCF between pairs of MGs and tracers (blue) after correction of the IC. The ACF and CCF are fitted by a power-law with $\beta=-0.8$, shown by dashed and solid lines, where the blue shaded region corresponds to the 1$\sigma$ error around the best fit power-law to the CCF. The error in the ACF fit, $\pm 0.045$, is not shown. The fitting range $40\arcsec<\theta<2000\arcsec$ is shown by yellow shade. (A colour version of this figure is available in the online journal.)}
    \label{fig:ACF_CCF}
\end{figure}

\begin{figure}
	\includegraphics[width=\columnwidth]{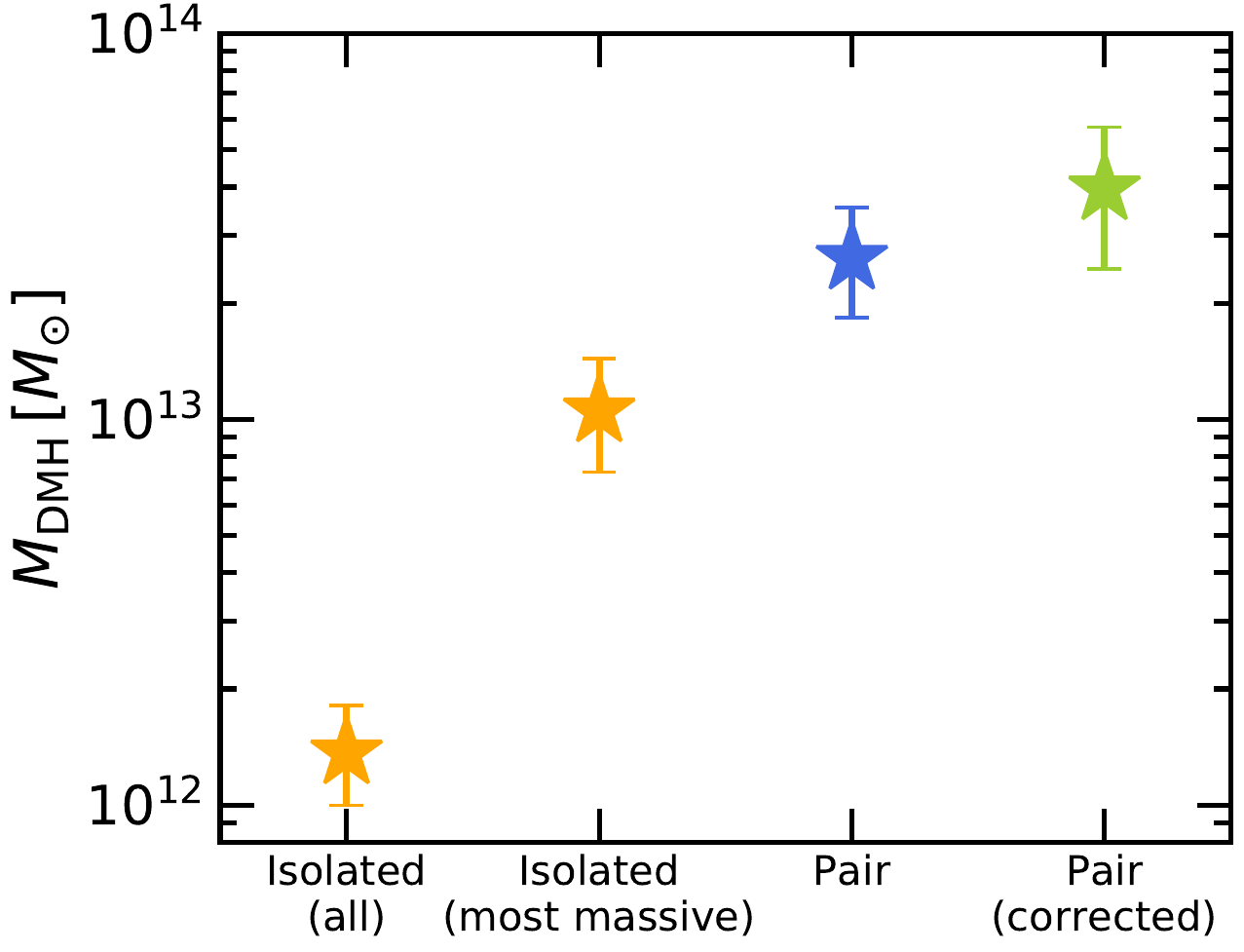}
    \caption{The mass of DMHs estimated by clustering analysis. A blue star indicates pairs of MGs. For comparison, the DMH masses of isolated MGs are also plotted. ``Isolated (all)" and ``isolated (most massive)" (orange) refer to non-pair galaxies whose stellar masses are larger than $10^{11}\, M_{\sun}$ and $10^{11.3}\, M_{\sun}$, respectively. In addition, we show the DMH mass of ``true pairs" assuming that the fraction of true pairs is 54\% (green) as calculated in Section~\ref{sec:true_pair_fraction}. (A colour version of this figure is available in the online journal.)}
    \label{fig:Mdh}
\end{figure}

Fig.~\ref{fig:ACF_CCF} shows the ACF and the CCF thus obtained. A signal is clearly detected for both correlation functions. From these correlation functions, we estimate the average DMH mass of the core candidates; we also estimate the average DMH masses of isolated (i.e. non-pair) MGs in a similar manner (Fig.~\ref{fig:Mdh}). We confirm that the core candidates are hosted by very massive haloes with $M_\mathrm{DMH}=2.6_{-0.8}^{+0.9}\times 10^{13}\, M_{\sun}$, which is within our target mass range. We also find that this value is larger than the DMH masses of isolated MGs with $\log(M_{*}/M_{\sun})\geq 11.0$ and $\log(M_{*}/M_{\sun})\geq 11.3$ by 1.3 dex and 0.4 dex, respectively, indicating that pairs of MGs can trace more massive haloes than their isolated counterparts.

\subsection{The fraction of true pairs and the intrinsic DMH mass}
\label{sec:true_pair_fraction}
Since we use a photo-\textit{z} galaxy catalogue, the detected pairs of MGs may be contaminated by false pairs due to chance projection. Although we cannot tell which pairs are true systems without spectroscopic follow up observation, we can statistically estimate the fraction of ``true pairs". Following the method introduced in \citet{Bethermin2014}, we estimate this fraction as a function of the maximum angular separation using the ACF of the MGs. 

In general, the ACF of galaxies is expressed by the sum of two components, the one-halo term and the two-halo term:
\begin{equation}
    \omega_\mathrm{ACF}(\theta) = \omega_\mathrm{1h}(\theta) + \omega_\mathrm{2h}(\theta),
\end{equation}
where $\omega_\mathrm{1h}$ and $\omega_\mathrm{2h}$ are the one-halo and two-halo terms, respectively. The one-halo term comes from galaxy pairs hosted by the same haloes and the two-halo term originates from pairs hosted by different haloes. Therefore, we can estimate the fraction of true pairs by evaluating the relative strength of the one-halo term. The fraction of true pairs whose separation is less than $\theta$ can be calculated as:
\begin{align}
\label{f_true}
    f_\mathrm{true}(\theta) &= \frac{\int_{0}^{\theta}\omega_\mathrm{1h}(\theta^{\prime})\theta^{\prime} d\theta^{\prime}} {\int_{0}^{\theta} \left[1+\omega_\mathrm{ACF}(\theta^{\prime})\right] \theta^{\prime} d\theta^{\prime}}.
\end{align}
We first calculate the ACF of the MGs. Then, we derive the two-halo term assuming that this term can be described as:
\begin{equation}
    \omega_\mathrm{2h}(\theta) = b^{2} \omega_\mathrm{DM}(\theta),
\end{equation}
where $b$, the normalisation, is the bias parameter and $\omega_\mathrm{DM}(\theta)$ is the angular ACF of dark matter calculated from the matter power spectrum. We fit $\omega_\mathrm{2h}(\theta)$ over $40\arcsec-2000\arcsec$ with correction of the IC. Finally, we use Equation~\eqref{f_true} to derive $f_\mathrm{true}$. Here we consider an additional correction of $\omega$. The ACF signal becomes weaker when the redshift window becomes larger. While the redshift window is 0.24 in our pair finder algorithm, that in this analysis is 1.5 ($1.5<z<3.0$). To correct for this effect, we multiply $\omega$ by 4.78, the typical ratio of $\omega_\mathrm{ACF}(\Delta z = 0.24)$ to $\omega_\mathrm{ACF}(1.5<z<3.0)$.

\begin{figure}
	\includegraphics[width=\columnwidth]{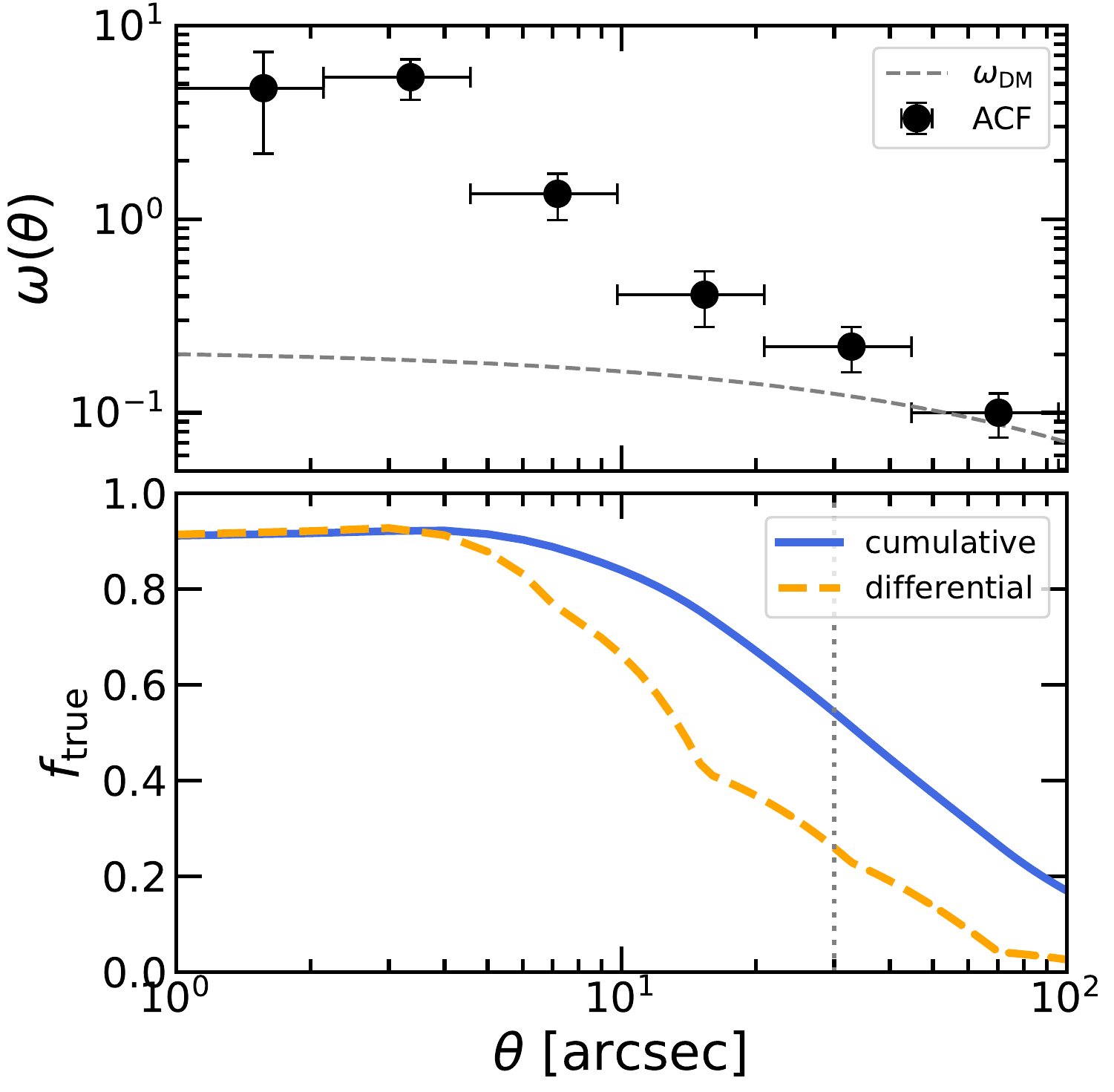}
    \caption{\textit{Top panel}: The ACF of MGs. Black points show the observed ACF. A dashed line shows the two-halo term of the ACF, derived from fitting in $40\arcsec< \theta <2000\arcsec$. \textit{Bottom panel}: The fraction of true pairs as a function of pair separation. The blue solid line shows $f_\mathrm{true}$ for pairs whose separations are smaller than $\theta$, which is given by Equation~\eqref{f_true}, while the orange dashed line is $f_\mathrm{true}$ at a given $\theta$. (A colour version of this figure is available in the online journal.)}
    \label{fig:f_true}
\end{figure}

Fig.~\ref{fig:f_true} shows the ACF of the MGs and $f_\mathrm{true}$. In our pair finder we adopt $30\arcsec$ as the maximum separation, resulting in $f_\mathrm{true}=54\%$.

Since isolated MGs have a weaker clustering signal than real pairs, the contamination by false pairs reduces the clustering signal of pairs. We estimate the bias of real pairs, $b_\mathrm{true}$, and hence the intrinsic DMH mass of cores with the following relation \citep{Bethermin2014}:
\begin{equation}
    b_\mathrm{pair}^{2} = f_\mathrm{true}b_\mathrm{true}^{2} + 2(1-f_\mathrm{true}b_\mathrm{c}^{2}),
\end{equation}
where $b_\mathrm{pair}$ is the bias parameter of the core candidates obtained in Section~\ref{sec:clustering} and $b_\mathrm{c}$ is the bias parameters of contaminants. We approximate $b_\mathrm{c}$ by the bias of the MGs. The intrinsic DMH mass is found to be $4.0_{-1.5}^{+1.8}\times 10^{13}\,M_{\sun}$, which is shown in Fig.~\ref{fig:Mdh} with label of ``corrected". 

Using the Millennium Simulation, \citet{Muldrew2015} have shown that
the most massive progenitor haloes at $z=2$ of present-day $M_\mathrm{DMH}=1\times 10^{14}\, M_{\sun}$ clusters have a median mass of $1.4\times 10^{13}\, M_{\sun}$, with a $1\sigma$
scatter of $0.22$ dex. The mean DMH mass of the cores exceeds this best-fitting median value even before contamination correction.

\begin{figure}
	\includegraphics[width=\columnwidth]{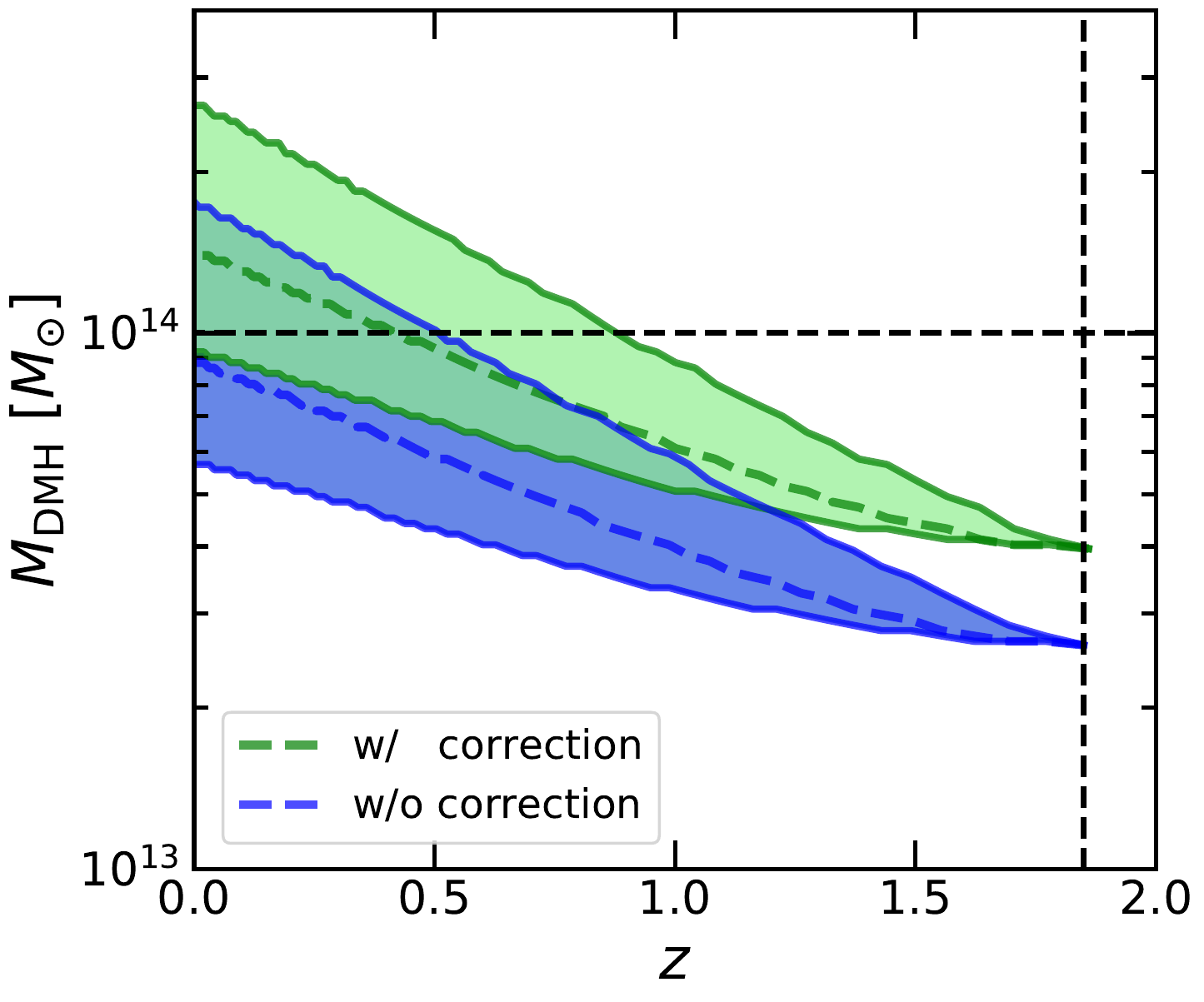}
    \caption{The descendant mass of the DMHs hosting pairs calculated by the extended Press-Schechter model from $z=1.85$ to $z=0$. Shaded regions show the $1 \sigma$ scatters and dashed lines are the modes. Blue and green colours correspond to initial masses of $2.62\times 10^{13}\, M_{\sun}$ and $3.96 \times 10^{13}\, M_{\sun}$, respectively. The latter is the intrinsic mass after correction of contamination (see Section~\ref{sec:true_pair_fraction}). (A colour version of this figure is available in the online journal.)}
    \label{fig:eps_descendant_halomass}
\end{figure}

Then we estimate the descendant DMH mass of the cores using the extended Press-Schechter model. We assume that all the cores are located at $z=1.85$. The descendant masses are shown in Fig.~\ref{fig:eps_descendant_halomass} as blue and green shades for masses before and after correction of contamination, respectively. We find that the host haloes of the cores can grow into $1\times 10^{14}\ M_{\sun}$ at $z=0$, comparable to the mass of a Virgo or Fornax-like cluster \citep{Chiang2013}.

\subsection{The number density of cores}
\label{sec:number_density}
To check whether our pair-finding method finds massive DMHs completely, we compare the number density of our core candidates to that derived from the halo mass function. Assuming that all of the most massive DMHs host a single pair of MGs, we first calculate the minimum mass of DMHs which host a pair ($M_\mathrm{min}$) as follows:

\begin{equation}
    b_\mathrm{obs} = \frac{ \int^{\infty}_{M_\mathrm{min}} b(M)\, \frac{dn(M)}{dM}\, dM }{ \int^{\infty}_{M_\mathrm{min}} \frac{dn(M)}{dM}\, dM },
\end{equation}
where $\frac{dn(M)}{dM}$ is the halo mass function and $b(M)$ is the bias parameter as a function of halo mass. Here, we adopt \citet{Sheth1999} as the halo mass function. Then we calculate the number density of DMHs which are more massive than $M_\mathrm{min}$. 

In Table~\ref{tab:number_density}, we summarise the number density of each population. Our core candidates have a lower number density than that estimated from the halo mass function by factor of 2.5 (3.5) with (without) true pair correction, resulting in $\sim 40\% \ (30\%)$ completeness. We further explore the completeness as a function of DMH mass in the next section using the IllustrisTNG simulation.

\begin{table}
\centering
    \caption{The number density of cores, DMHs and clusters.}
    \label{tab:number_density}
	\begin{tabular}{lccc}
		\hline
		objects & redshift & $M_\mathrm{min}\,[M_{\sun}]$ &  $n\,[\mathrm{cMpc^{-3}}]$ \\
		\hline
		cores & $1.5\leq z\leq 3.0$ & $1.6\times 10^{13}$ & $2.8 \times 10^{-6}$  \\  [2pt]
		DMHs$^a$ & $1.5\leq z\leq 3.0$ & $1.6\times 10^{13}$ & $1.0\times 10^{-5}$  \\  [2pt]
		cores (true)$^b$\ & $1.5\leq z\leq 3.0$ & $2.5\times 10^{13}$ & $1.5\times 10^{-6}$  \\  [2pt]
		DMHs$^a$ & $1.5\leq z\leq 3.0$ & $2.5\times 10^{13}$ & $4.0\times 10^{-6}$  \\  [2pt]
		local clusters & $z=0$ & $1.0\times 10^{14}$ & $1.5\times 10^{-5}$ \\
		\hline
	\end{tabular}
	\begin{tablenotes}[normal]
	 \item \textit{Notes.} $^a$The number density of DMHs calculated by the halo mass function. $^b$A true pair fraction of $54\%$ is considered.
    \end{tablenotes}
\end{table}

\subsection{Comparison with the IllustrisTNG}
\label{sec:illustris}
In this paper, we assume that pairs of MGs are typically hosted by more massive DMHs than isolated MGs. In the previous section, we confirm this hypothesis in a statistical manner with observational data. However, as shown in Section~\ref{sec:number_density}, our method may not be able to find all massive DMHs. To evaluate the effectiveness of pairs as tracers of cores, we need to know the mass distribution of pair-host DMHs, the fraction of massive DMHs which host pairs, and the fraction of pair-host DMHs which can actually grow into $M_\mathrm{DMH}\geq 10^{14}\, M_{\sun}$ at $z=0$. Since observational data do not tell us individual halo masses, we employ a mock galaxy catalogue of the IllustrisTNG project for this purpose.

The IllustrisTNG project is a series of cosmological magnetohydrodynamical simulations of galaxy formation and evolution including various baryon physics: star formation, stellar evolution, chemical enrichment, primordial and metal-line cooling of the gas, stellar feedback, and black hole formation, growth and feedback \citep{Pillepich2018a,Weinberger2017}. The simulations consist of three runs with different box sizes and each run also has three different resolutions. We use results from TNG300 which has the largest volume $\sim( 205\, \mathrm{cMpc}/h)^3$ among the three runs. Thanks to the large volume, TNG300 is suitable for investigating properties of rare objects like galaxy clusters. Among the three TNG300 runs, we select the one with the highest mass resolution, TNG300-1, and use the halo (group) and galaxy (subhalo) catalogues as well as merger trees (i.e. the merger histories of individual haloes). A detailed description about the simulations is found in IllustisTNG presentation papers \citep{Naiman2018,Springel2018,Pillepich2018c,Marinacci2018,Nelson2018}.

First, from the mock galaxy catalogue of $z=2$ (snapshot 33), we extract the positions and stellar masses of galaxies. Then we select galaxies with $\log(M_{*}/M_{\sun})\geq 11$ and apply the pair finder to them. Instead of the angular separation criterion in Section~\ref{sec:pair_finder}, we consider a condition that three dimensional separations are $< 0.3\, \mathrm{pMpc}$. We identify 103 pairs from 2092 massive mock galaxies. The number of independent pair-host haloes is 100 because some pairs are hosted by the same haloes.

\begin{figure*}
	\includegraphics[width=2\columnwidth]{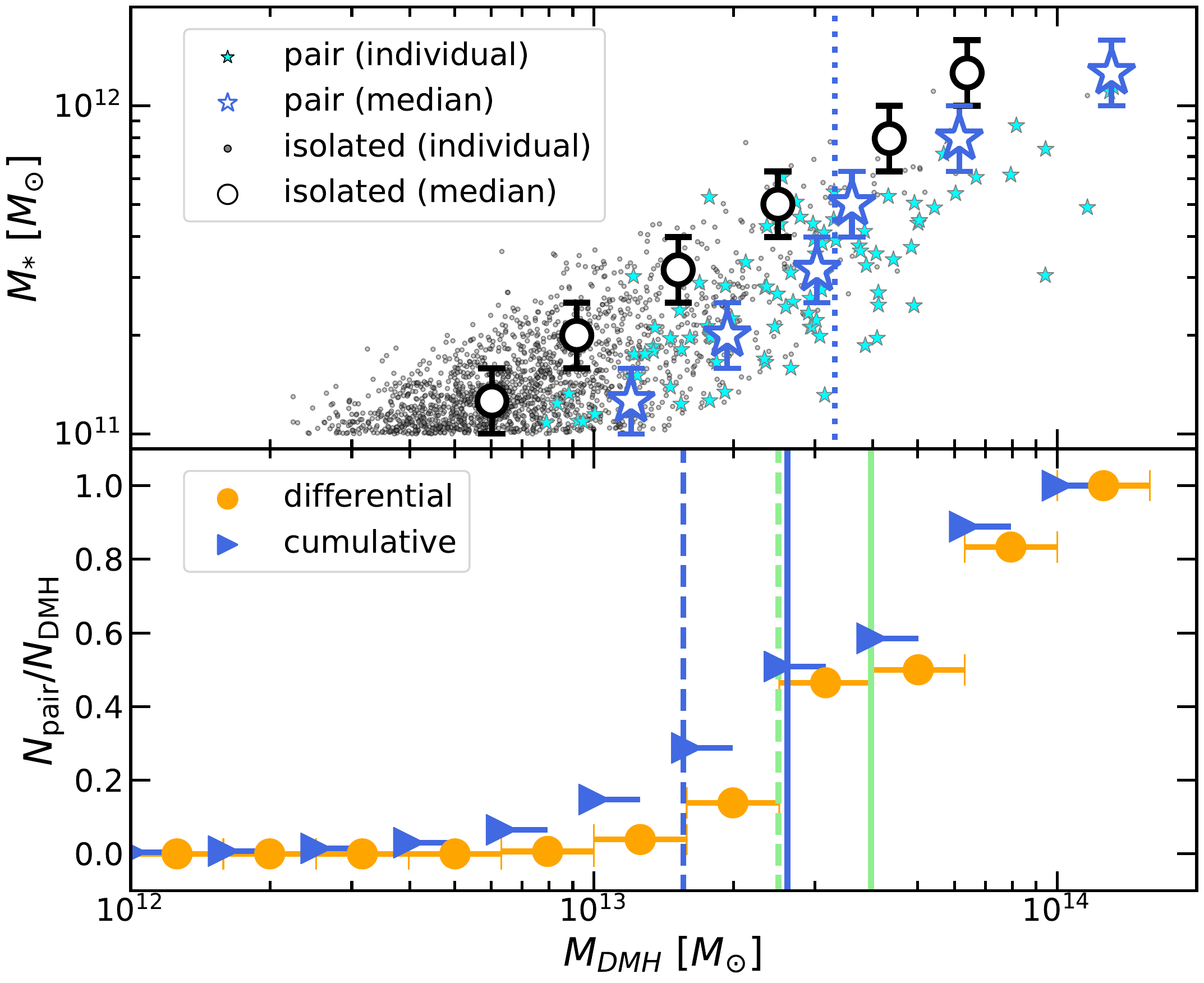}
    \caption{\textit{Top panel}: The relation between stellar mass and host DMH mass for massive galaxies ($\log(M_{*}/M_{\sun})\geq 11$) in the IllustrisTNG300-1. Small cyan stars refer to pairs of massive galaxies with separations smaller than $0.3\, \mathrm{pMpc}$ while dots show isolated central galaxies. Large stars and circles show corresponding median values in 0.2 dex stellar mass bins. A blue dotted line is the average $M_\mathrm{DMH}$ of pair-host DMHs in IllustrisTNG300-1. \textit{Bottom panel}: The fraction of DMHs which host pairs of massive galaxies as a function of halo mass. Blue triangles show the pair-host fraction of DMHs more massive than a given mass and orange circles represent the differential fraction. Solid and dashed lines show the average $M_\mathrm{DMH}$ of the pairs estimated by clustering analysis and $M_\mathrm{min}$ obtained in Section~\ref{sec:number_density}, where blue and green colours show, respectively, before and after true pair correction. (A colour version of this figure is available in the online journal.)}
    \label{fig:Illustris-pair}
\end{figure*}

In the top panel of Fig.~\ref{fig:Illustris-pair} we show the relation between the stellar masses of central galaxies and their host DMH masses\footnote{We approximate DMH masses by $M_{200}$, which represents the total mass enclosed by a sphere whose inner mass density is 200 times the critical density of the universe.}. Small stars and dots mean DMHs which host pairs and isolated centrals, respectively. For pair-host DMHs, we plot the largest stellar mass among each pair. For a series of stellar mass bins with a width of 0.2 dex, we calculate median DMH masses. Large stars and circles show the median masses of pair and isolated central host DMHs, respectively. We find that at a fixed stellar mass, the median mass of DMHs which host a pair is larger by 0.15 to 0.3 dex. This suggests that pairs of MGs are effective tracers of the most massive DMHs in the universe at $z\sim2$.

We also show the fraction of DMHs which host a pair as a function of halo mass in the bottom panel of Fig.~\ref{fig:Illustris-pair}. Blue triangles show the pair-host fraction of DMHs which is more massive than a given mass and orange circles represent the differential fraction. DMH masses estimated from clustering analysis and $M_\mathrm{min}$ obtained in Section~\ref{sec:number_density} are also plotted as solid and dashed lines, respectively. At the mass of $M_\mathrm{min}$ with (without) true pair correction, the cumulative pair-host fraction is $\sim 50\%\ (30\%)$, being consistent with the completeness calculated from the halo mass function. Furthermore, the pair-host fraction monotonically increases with halo mass. These results mean that pairs of MGs can effectively trace DMHs which are massive enough to be regarded as proto-cluster cores.

Finally, we investigate the fraction of pair-host haloes at $z=2$ that can evolve into $\geq 10^{14}\, M_{\sun}$ at $z=0$. Tracing merger histories of pair-host haloes, we find that 100 independent pair-host haloes at $z=2$ become 89 independent haloes at $z=0$, indicating that mergers reduce $\sim 10\%$ of pair-host haloes. Among those descendants, 63 haloes are more massive than $10^{14}\, M_{\sun}$, which are regarded as clusters. This means that the purity of pair-host haloes as tracers of proto-cluster cores is $63\%$. In the simulation box, there are 280 clusters. Therefore, the completeness of pairs as tracers of $z=0$ clusters is $23\%$. We further investigate the completeness for $z=0$ clusters in terms of their mass. Following \citet{Chiang2014}, we divide $z=0$ clusters into three types according to their mass: Fornax-like ($M_\mathrm{DMH}=1-3 \times 10^{14}\,M_{\sun}$), Virgo-like ($M_\mathrm{DMH}=3-10 \times 10^{14}\,M_{\sun}$) and Coma-like ($M_\mathrm{DMH}>1 \times 10^{15}\,M_{\sun}$) clusters. At $z=0$, the numbers of descendants of pair-host haloes (and all haloes in the simulation box) classified as Fornax-like, Virgo-like and Coma-like clusters are 38 (235), 22 (42), 3 (3), respectively, resulting in $16\%$, $52\%$ and $100\%$ completeness for each type. This suggests that pairs of MGs are not only good tracers of the progenitor haloes of the most massive clusters but also those of Virgo-like clusters.

\begin{figure}
	\includegraphics[width=\columnwidth]{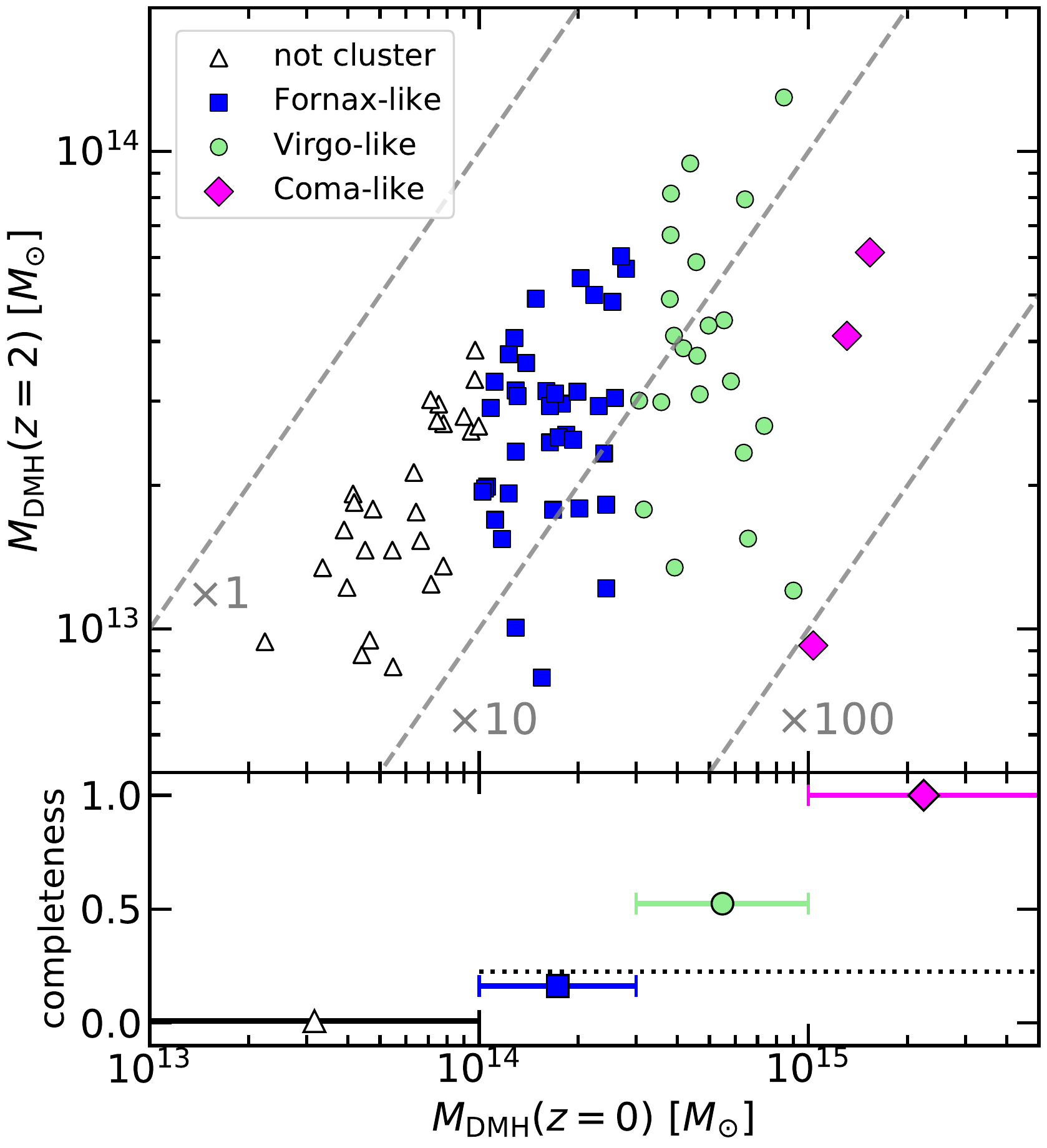}
    \caption{\textit{Top panel}: The DMH masses of pair-host haloes and those of the descendants at $z=0$. Blue squares, green circles and magenta diamonds show the masses of progenitor haloes of Fornax-like, Virgo-like and Coma-like clusters, respectively. Progenitors which have less massive descendants than $10^{14}\, M_{\sun}$ are shown by black triangles. Grey dashed lines show the ratio of $M_\mathrm{DMH}(z=0)$ to $M_\mathrm{DMH}(z=2)$. \textit{Bottom panel}: The completeness of pairs of massive galaxies as tracers of $z=0$ clusters in four mass bins. The meanings of the symbols are the same as those in the top panel. The black dotted line shows the completeness in the whole mass range above $1\times10^{14}\,M_{\sun}$, which is 0.23. (A colour version of this figure is available in the online journal.)}
    \label{fig:Illustris-descendant}
\end{figure}

In Figure~\ref{fig:Illustris-descendant}, we show the DMH masses of pair-host haloes and their descendants at $z=0$. At fixed $M_\mathrm{DMH}(z=0)$, the masses of progenitors have a 1$\sigma$ scatter of $0.2-0.4$ dex, which is similar to the value found by \citet{Muldrew2015}. This relatively large scatter implies that there are various paths of halo mass growth. For each type of clusters, we check the fraction of DMHs which become more than ten times more massive from $z=2$ to $z=0$. For Fornax-like, Virgo-like and Coma-like clusters, these fractions are roughly 15\%, 60\% and 100\%, respectively, suggesting that the progenitors of more massive clusters tend to grow more rapidly after $z=2$.

\section{Properties of member galaxies of proto-cluster cores}
We examine the stellar mass function (SMF) and the quiescent fraction for galaxies in the detected cores. Since the COSMOS2015 catalogue is a photo-\textit{z} sample, we subtract field galaxies statistically as described below.

\subsection{Field subtraction and the field stellar mass function}
We extract all galaxies down to $\log(M_{*}/M_{\sun})=9.0$ in cylindrical regions around the 75 cores with a radius of $\Delta r=0.3\,\mathrm{pMpc}$ and a line of sight length $\Delta z=0.5$. We adopt this relatively large $\Delta z$ value not to miss low-mass galaxies near the mass limit that have much larger photo-\textit{z} uncertainties than $\log(M_{*}/M_{\sun})\geq 11.0$ galaxies. The galaxies in these cylindrical regions are contaminated by field galaxies. We perform field subtraction in the following manner.

First, we calculate the SMFs of field galaxies by dividing the galaxy sample of $\log(M_{*}/M_{\sun})\geq 9.0$ into 20 redshift bins of range $1.25<z<3.25$ and width $\Delta z=0.1$. For each redshift bin, we also compute the total cosmic volume occupied by the cylindrical regions around the cores. Then, multiplying the field SMFs by these cosmic volumes, we estimate the total number of contamination galaxies falling within the 75 cylindrical regions as a function of stellar mass. Finally, we subtract this mass function of contaminants from the raw counts around the cores.

We also need a field SMF averaged over $1.5<z<3.0$ that is compared with the SMF of member galaxies. Because the redshift distribution of the core sample is slightly different from that of the general galaxy sample, we calculate this field SMF as:

\begin{equation}
\label{phi_field}
    \Phi_\mathrm{field} = \frac{\sum_{i}{n(z_{i})\Phi_{\mathrm{field},\, i}}}{\sum_{i}{n(z_{i})}},
\end{equation}
where ${z_{i}}$ is the i-th redshift bin, $n(z_{i})$ is the number of cores at ${z_{i}}$, and $\Phi_{\mathrm{field},\,i}$ is the field SMF at $z_{i}$.

\subsection{The stellar mass function}
\begin{figure*}
	\includegraphics[width=2\columnwidth]{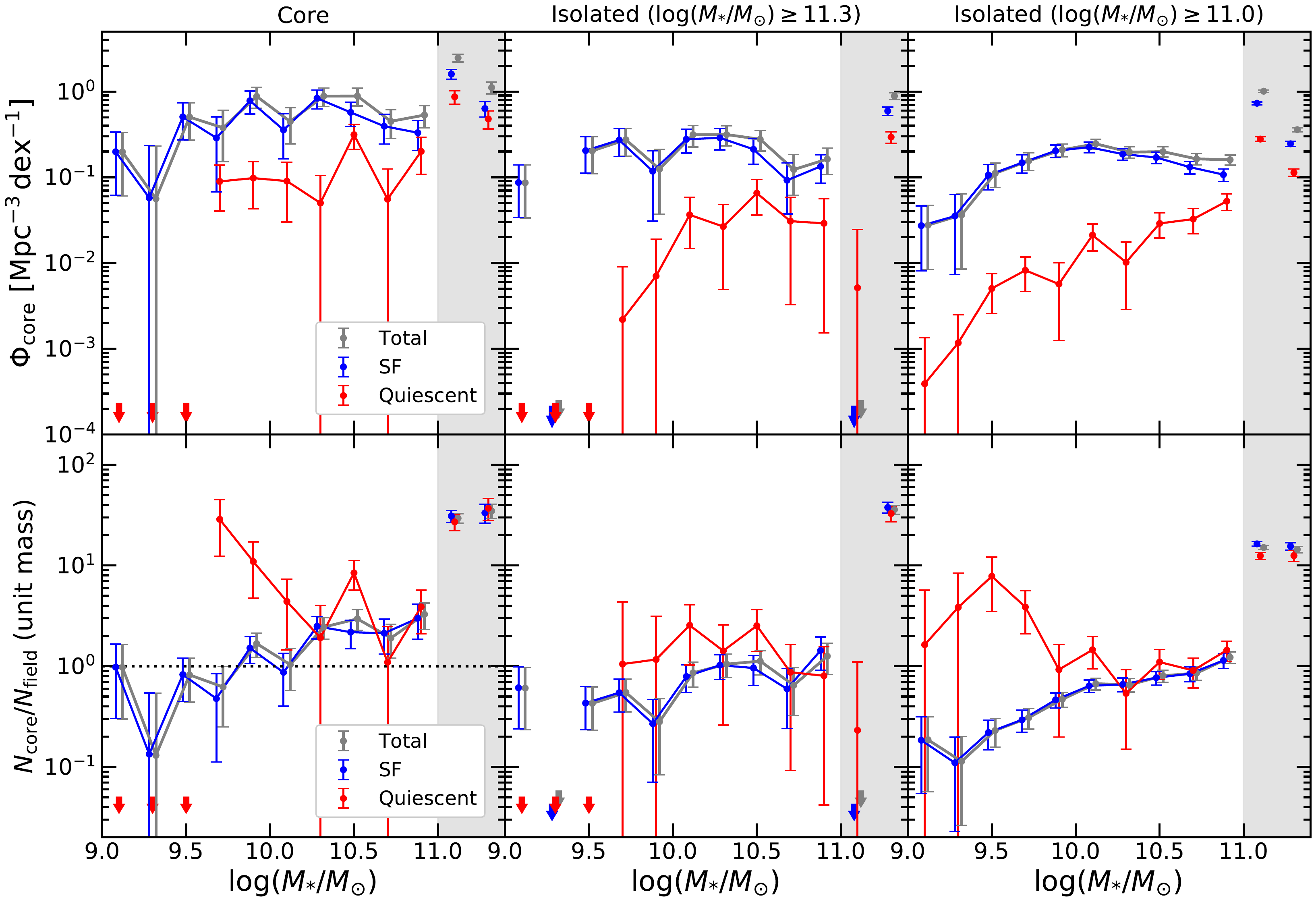}
    \caption{\textit{Top panel}: The stellar mass functions (SMFs) of galaxies in the cores (left), around the most massive ($\log(M_{*}/M_{\sun})\geq 11.3$) isolated galaxies (middle), and around massive ($\log(M_{*}/M_{\sun})\geq 11.0$) isolated galaxies (right). Detection incompleteness has not been corrected. Grey, blue and red colours mean the SMFs of all galaxies, star-forming galaxies, and quiescent galaxies, respectively. Grey shaded regions show the mass range suffering from selection bias. \textit{Bottom panel}: Same as top panels but divided by the field SMFs and normalised by total mass using Equation~\eqref{SMF_unitmass}. A black dotted line indicates unity. This normalisation is only valid for the cores. Arrows mean a negative value. (A colour version of this figure is available in the online journal.)}
    \label{fig:SMF}
\end{figure*}

The SMFs of galaxies in the cores and that of the field galaxies are shown in the top panel of Fig.~\ref{fig:SMF}. To calculate the former, we assume that DMHs hosting a pair are spheres with a radius of $0.3\,\mathrm{pMpc}$. Completeness correction as a function of stellar mass is not considered. In Fig.~\ref{fig:SMF}, grey, blue and red lines refer to the SMFs of total galaxies, star-forming galaxies and quiescent galaxies, respectively. For comparison, we calculate the SMFs around isolated MGs with stellar masses of $\log(M_{*}/M_{\sun})\geq 11.3$ and $\log(M_{*}/M_{\sun})\geq 11.0$ in the same way as that for pairs. Note that above $\log(M_{*}/M_{\sun})=11$, the SMFs are positively biased because there are at least two (one) MGs in each core (each isolated MG) which are used to identify them.

It is found that the SMFs of total and star-forming galaxies in the cores as well as around isolated MGs have a flat shape below $\log(M_{*}/M_{\sun})=11$, where the SMFs are not affected by selection bias. We also find that the normalisation of the SMF of the cores is roughly twice as large as those of the two classes of isolated MGs, meaning that the pairs reside in denser environment.

To discuss the shapes of the SMFs and the galaxy formation efficiency in the cores, we also calculate the ratio between the SMF of member galaxies and that of field galaxies for each star formation class. The normalisations of the SMFs of the cores are roughly two to three orders of magnitude higher than those of the field galaxies. We again normalise the ratio of SMFs by total mass as:
\begin{equation}
\label{SMF_unitmass}
    \frac{N_\mathrm{core}}{N_\mathrm{field}} = \frac{\Phi_\mathrm{core}}{\Phi_\mathrm{field}} \frac{\rho_\mathrm{crit}\Omega_\mathrm{m}V_\mathrm{core}}{M_\mathrm{core}},
\end{equation}
where $\rho_\mathrm{crit}$ is the critical density of the universe in our cosmology, $V_\mathrm{core}$ is the average comoving volume and $M_\mathrm{core}$ is the DMH mass of the cores, respectively.

The results are plotted in the bottom panel of Fig.~\ref{fig:SMF}. We find that this ratio increases with stellar mass. In other words, the member galaxies of proto-cluster cores have a more top-heavy SMF than field galaxies. This result is qualitatively consistent with the simulation \citep{Lovell2018,Muldrew2018}. We note that the SMFs of field galaxies are not exactly the same among the three panels because the redshift distributions $n({z_{i}})$ of corresponding massive galaxy populations are different. See the definition of $\Phi_\mathrm{field}$ in Equation~\eqref{phi_field}.

We also find that the ratio of the SMFs is below unity, although marginal, at $\log(M_{*}/M_{\sun})\lesssim 10$ and above unity at $\log(M_{*}/M_{\sun})\gtrsim 10$, meaning that in core regions, the formation of low-mass galaxies may be suppressed while that of high-mass galaxies is enhanced compared to the field. Destruction of low-mass galaxies by mergers and/or tidal disruption \citep{Martel2012} are possible causes of the lower formation efficiency of low-mass galaxies. Another possibility is the suppression of star formation of low-mass galaxies. As described in Section~\ref{sec:fq}, low-mass galaxies in the cores have a higher quiescent fraction than their field counterparts. This may support this possibility. For high-mass galaxies, the high density environment of proto-cluster cores may enhance the formation of high-mass galaxies by the early formation of large DMHs and/or more frequent mergers \citep{Muldrew2018}.

Trends similar to those seen in the SMFs of the cores have been found in several observational studies which focus on both global and local environments. At high redshift, $z\sim 2.5$, \citet{Shimakawa2018} have found that the SMF of $\mathrm{H \alpha}$ emitters in the densest regions of a proto-cluster is more top-heavy than that in less dense regions in terms of a clear excess of high-mass galaxies ($\log(M_{*}/M_{\sun})>10.5$), although they have not been able to find a clear difference at the low-mass end. Together with the evidence that high-mass galaxies in the densest regions are more actively star-forming, they have concluded that the formation of massive galaxies has been accelerated in the densest parts of a proto-cluster. Our SMF for star-forming galaxies in the cores is qualitatively consistent with these results, implying an enhancement of the star formation of high-mass galaxies in the cores.

Differences in the SMFs between mature clusters and fields have also been reported at $z\lesssim 1.5$. \citet{VanderBurg2013,VanderBurg2018} have shown that cluster galaxies have more top-heavy SMFs at $0.5<z<1$ primarily because of a shallower low-mass end slope, especially for quiescent galaxies. \citet{Nantais2016} have reported that in clusters at $z\sim 1.5$, the SMF of quiescent galaxies with low stellar masses ($\log(M_{*}/M_{\sun})\lesssim10.5$) has a roughly 50\% contribution to the total SMF, while only 20\% in fields. They interpret this as environmental quenching of low-mass galaxies, although they do not find a clear difference in the shape of the SMF of total galaxies between clusters and fields. At $\log(M_{*}/M_{\sun})\sim 10$, our SMF of quiescent galaxies in the cores shows higher $\Phi_\mathrm{core}/\Phi_\mathrm{field}$ values compared to more massive bins. This may imply that cores at $z\sim 2$ are similar to mature clusters $z\lesssim 1.5$ in terms of a higher fraction of low-mass quiescent galaxies than fields. We should note that the SMF of quiescent galaxies has negative values at the lowest-mass bins ($\log(M_{*}/M_{\sun})\lesssim 9.5$) possibly due to low statistics.

The effect of local environment on galaxy formation has been studied by many papers. Using the Bayesian motivated N-th nearest neighbour as an environment measure \citep{Cowan2008}, \citet{Kawinwanichakij2017} have shown that quiescent galaxies are likely to reside in denser environments than star-forming ones even at fixed stellar mass at $0.5<z<2.0$. The same trend has also been reported in \citet{Malavasi2016}, which have used the number density of galaxies within a cylindrical region as an environment measure. These results are qualitatively consistent with our results. At lower redshift ($0.55<z<1.3$), \citet{Tomczak2017} have found strong dependence of the shape of the SMFs on local environment. They have used Voronoi tessellation \citep{Darvish2015} as an environment measure and shown that galaxies in denser environments have more top-heavy SMFs than in fields for both star-forming and quiescent galaxies, which is similar to what we find. With the same environmental measure as \citet{Kawinwanichakij2017}, \citet{Papovich2018} have argued that there are not major differences in the shape of the SMF for either star-forming or quiescent galaxies between high- and low-density environments at $1.5<z<2.0$. However, they have also pointed out that the SMFs of star-forming galaxies at $\log(M_{*}/M_{\sun})\sim 10.5$ in dense environments show an excess, which is seen in our SMF. These past studies are broadly consistent with ours when the differences in the definition of local environment are taken into account. Note that proto-cluster cores are extremely high-density regions where the 3D galaxy density is two orders of magnitude higher than the cosmic average as found in the comparison of the SMFs between cores and fields. In any case, proto-cluster cores are the most promising places to detect environmental dependence in the early universe.

\subsection{The quiescent fraction}
\label{sec:fq}
\begin{figure*}
	\includegraphics[width=2\columnwidth]{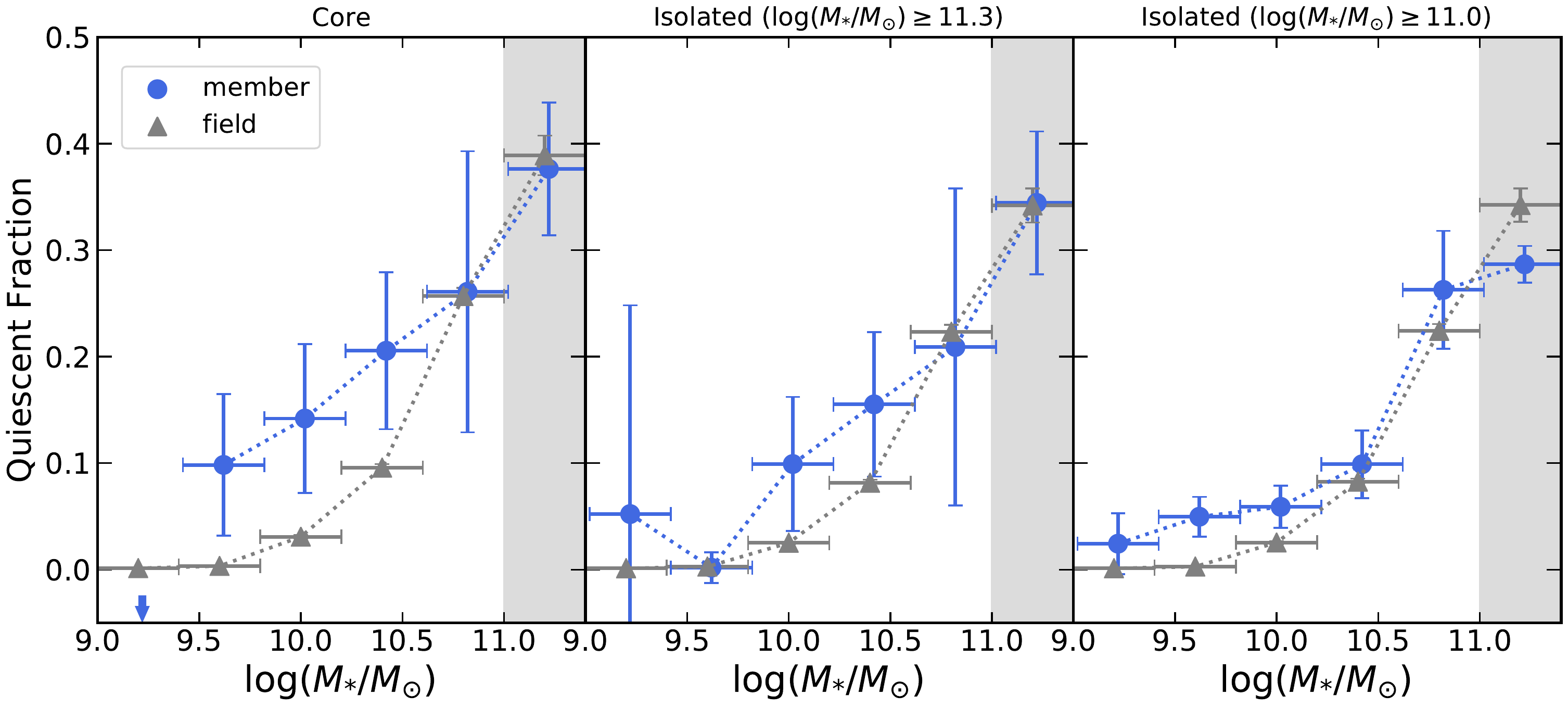}
    \caption{The quiescent fraction ($f_\mathrm{q}$) in the cores (left), around the most massive ($\log(M_{*}/M_{\sun})\geq 11.3$) isolated galaxies (middle), and around massive ($\log(M_{*}/M_{\sun})\geq 11.0$) isolated galaxies (right) plotted as blue symbols. The $f_\mathrm{q}$ of field galaxies is also plotted in each panel (grey symbols). In the mass range of $\log(M_{*}/M_{\sun})>11$, which is coloured in grey, $f_\mathrm{q}$ is affected by selection bias. An arrow means $f_\mathrm{q}<0$ due to field subtraction. (A colour version of this figure is available in the online journal.)}
    \label{fig:fq}
\end{figure*}

We measure the quiescent fraction for galaxies in the cores. Here, the quiescent fraction $f_\mathrm{q}$ is defined as\\
\begin{equation}
    f_\mathrm{q}=\frac{N_\mathrm{q}}{N_\mathrm{total}},
\end{equation}
where $N_\mathrm{total}$ and $N_\mathrm{q}$ are the numbers of total and quiescent galaxies, respectively. As in the previous section, we also compute $f_\mathrm{q}$ for galaxies around the two classes of isolated MGs for comparison. The results are shown in Fig.~\ref{fig:fq}.

It is found that all three environments have a higher quiescent fraction than the field. In each panel, the quiescent fraction of member galaxies is higher than in the field at $\log(M_{*}/M_{\sun})\lesssim 10.6$ while it is almost the same at $\log(M_{*}/M_{\sun})\gtrsim 10.6$. This probably reflects the fact that satellite galaxies around massive centrals are more likely to be quenched than isolated galaxies even at $z\sim 2$ \citep{Kawinwanichakij2016,Ji2018}. Interestingly, the $f_\mathrm{q}$ in the cores is higher than those in the others. In Table \ref{tab:fq_wholemass}, we summarise $f_\mathrm{q}$ in the whole mass range below $10^{11}\, M_{\sun}$, where galaxy number counts are not directly affected by selection bias. 
The $f_\mathrm{q}$ of galaxies in the cores is $17_{-4}^{+4}\%$, which is $3.3_{-0.8}^{+0.8}$ times higher than that of field galaxies, while that of galaxies around isolated MGs with $\log(M_{*}/M_{\sun})\geq11.0\,(11.3)$ is $11_{-2(4)}^{+2(4)}\%$, which is $2.4_{-0.4\,(0.8)}^{+0.4\,(0.8)}$ times higher than field galaxies. This suggests that proto-cluster cores are more evolved systems than DMHs hosting isolated MGs.

The value of $f_\mathrm{q}$ has been examined for several individual clusters at $1.6<z<1.8$, and much higher values than what we find have been reported: $f_\mathrm{q}\gtrsim 30\%$ at $\log(M_{*}/M_{\sun})\lesssim10.5$ and $f_\mathrm{q}\gtrsim 80\%$ at $\log(M_{*}/M_{\sun})\gtrsim10.5$ (\citealp{Newman2014,Cooke2016,Lee-Brown2017}). These differences may partly come from the fact that the clusters in these previous studies are more massive ($M_\mathrm{DMH}\gtrsim 8\times10^{13}\, M_{\sun}$) and thus more evolved systems than our cores. Part of the differences may also be due to cluster-to-cluster variation because these studies are each based on only a single cluster.

Then, we also calculate the environmental quenching efficiency (QE):
\begin{equation}
    QE=\frac{f_\mathrm{q}^\mathrm{member}-f_\mathrm{q}^\mathrm{field}}{1-f_\mathrm{q}^\mathrm{field}} = 1-\frac{f_\mathrm{sf}^\mathrm{member}}{f_\mathrm{sf}^\mathrm{field}},
\end{equation}
where $f_\mathrm{q}^\mathrm{member}$ and $f_\mathrm{q}^\mathrm{field}$ ($f_\mathrm{sf}^\mathrm{member}$ and $f_\mathrm{sf}^\mathrm{field}$) are the quiescent (star-forming) fraction of galaxies in the environment in question and in the field. This quantity describes what fraction of star-forming galaxies in the field would be additionally quenched if they were in the given environment.
The QE for the cores, $0.13_{-0.04}^{+0.04}$, is higher than that for the isolated MGs with $\log(M_{*}/M_{\sun})\geq11.0\,(11.3)$, $0.07_{-0.02\,(0.04)}^{+0.02\,(0.04)}$.

\begin{figure}
	\includegraphics[width=\columnwidth]{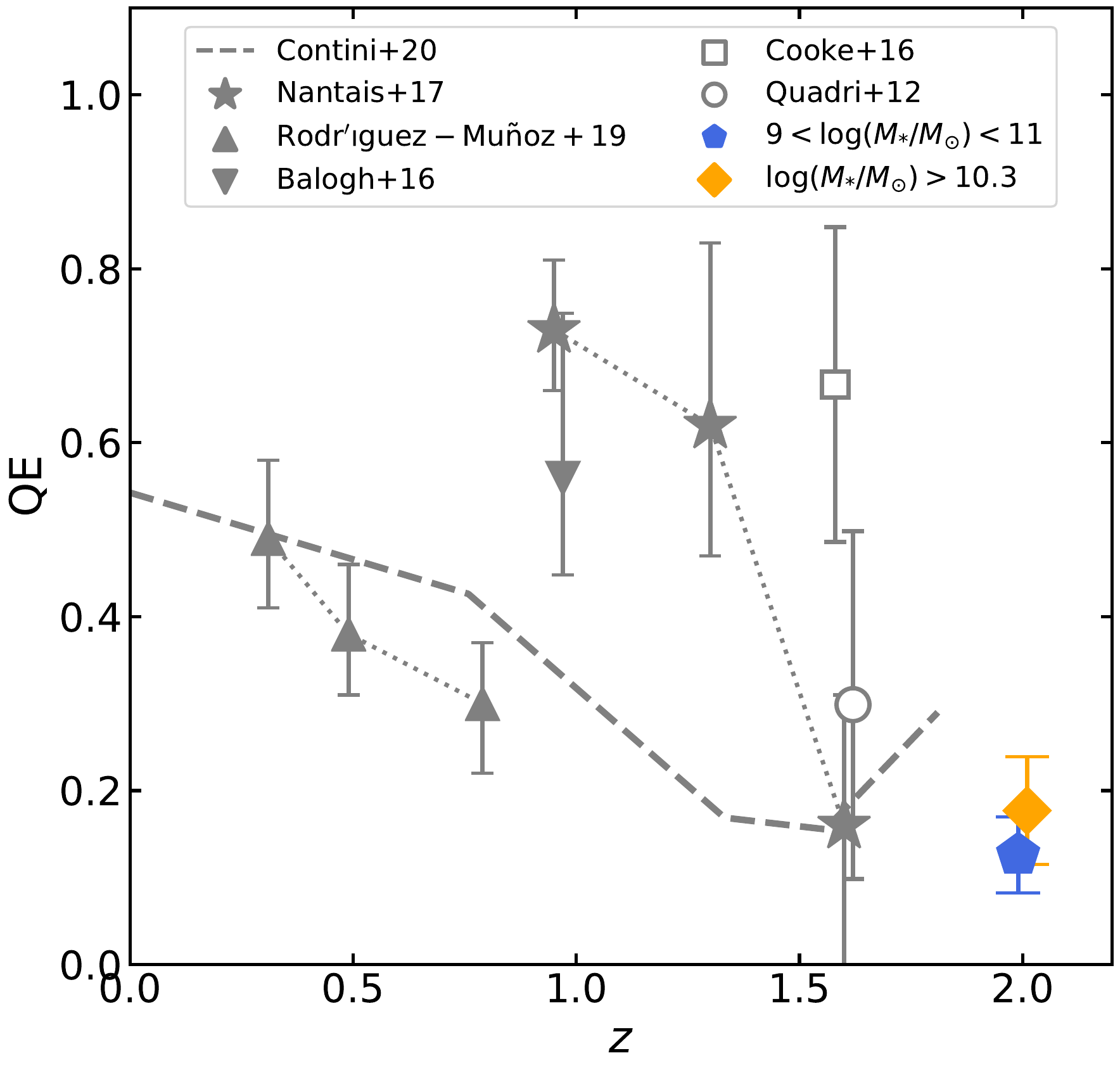}
    \caption{Environmental quenching efficiency (QE), defined as $QE=(f_\mathrm{q}^\mathrm{member}-f_\mathrm{q}^\mathrm{field})/(1-f_\mathrm{q}^\mathrm{field})$, as a function of redshift. A blue pentagon and an orange diamond are the QEs of the cores calculated for galaxies with $9.0<\log(M_{*}/M_{\sun})<11$ and $\log(M_{*}/M_{\sun})>10.3$, respectively. The other grey symbols are QEs in the literature. Stars, upward triangles, a downward triangle, an open circle and an open square are the QEs for cluster environments presented in \citet{Nantais2017,Rodriguez2019,Balogh2014,Cooke2016,Quadri2012}, respectively. Open symbols indicate QEs for individual clusters. A grey dashed line shows the QE calculated for clusters in an analytic galaxy formation model \citep{Contini2020}. (A colour version of this figure is available in the online journal.)}
    \label{fig:comp_QE}
\end{figure}

In Fig.~\ref{fig:comp_QE}, we plot the QE measurement of the cores (blue pentagon) together with those of known clusters in the literature. \citet{Quadri2012} and \citet{Cooke2016} have each calculated the QE for a single cluster at $z\sim 1.6$, using galaxies with $\log(M_{*}/M_{\sun})\geq10$ and $10\leq \log(M_{*}/M_{\sun})\leq 10.7$, respectively. \citet{Nantais2017}, \citet{Rodriguez2019} and \citet{Balogh2014} have measured QEs using 14, 24 and 10 clusters at various redshifts. \citet{Nantais2017} and \citet{Rodriguez2019} have used galaxies with $\log(M_{*}/M_{\sun})\geq10.3$ and $\log(M_{*}/M_{\sun})\geq10.0$, respectively. For the QE of \citet{Balogh2014}, we plot the result for $\log(M_{*}/M_{\sun})=10.5$. To classify galaxies into star-forming or quiescent, all the above studies have used a colour-colour diagram based on \citet{Williams2009}. \citet{Contini2020} have calculated the QE for clusters in an analytic galaxy formation model. They define clusters as DMHs with $\log(M_{*}/M_{\sun})>14.2$, and use galaxies with $\log(M_{*}/M_{\sun})\geq 9.5$. They define quiescent galaxies as those with a lower specific star formation rate than the inverse of the Hubble time. As a general trend, the QE becomes lower with increasing redshift. A qualitatively similar trend has been found for the QE of galaxies in locally dense environments \citep{Peng2010,Kawinwanichakij2017,Chartab2020}. One needs to be careful when comparing individual QE values directly, because the QE data in Fig.~\ref{fig:comp_QE} are heterogeneous in terms of the identification of clusters, the selection method of galaxies, and the stellar mass range used to calculate QEs.

For a detailed comparison, we focus on the result of \citet{Nantais2017} shown by grey stars. They have found that the QE changes dramatically after $z\sim 1.5$, from $QE\sim 0.16$ at $z\sim 1.6$ to $QE\sim 0.62$ at $z\sim 1.3$. To compare the QE for the cores with those obtained by \citet{Nantais2017}, we calculate it again by using galaxies in the mass range of $\log(M_{*}/M_{\sun})>10.3$ (orange diamond).
We find that the QE of the cores is positive, meaning that some mechanisms of environmental quenching have already worked in $z\sim 2$ cores. In addition, the QE of the cores is almost the same value as of mature clusters in \citet{Nantais2017} at $z\sim 1.6$ although the DMH mass of the cores is one-order of magnitude smaller than those of the $z\sim 1.6$ clusters. This result supports a scenario that cluster environments have not quenched galaxies significantly until $z\sim 1.5$ when a whole proto-cluster region starts to collapse, although excess quenching is already seen in cores. We note that at $z\sim 1.6$ the descendant mass of the cores does not reach $10^{14}\, M_{\sun}$, meaning that our cores may not be the progenitors of the $z\sim 1.6$ clusters. 

\begin{table}
	\centering
	\caption{The quiescent fraction ($f_\mathrm{q}$) and the environmental quenching efficiency (QE) of member galaxies in cores and around two classes of massive isolated galaxies, and those of corresponding field galaxies.}
	\label{tab:fq_wholemass}
	\begin{tabular}{lcccc}
		\hline
		objects & $f_\mathrm{q}^\mathrm{member}$ & $f_\mathrm{q}^\text{field}$ & $f_\mathrm{q}^\mathrm{member}$/$f_\mathrm{q}^\mathrm{field}$ & $QE$ \\
		\hline
		core & $0.17_{-0.04}^{+0.04}$ & $0.052_{-0.001}^{+0.001}$ & $3.3_{-0.8}^{+0.8}$ & $0.13_{-0.04}^{+0.04}$ \\  [6pt]
		iso\_{11.3}$^a$ & $0.11_{-0.04}^{+0.04}$ & $0.045_{-0.001}^{+0.001}$ & $2.4_{-0.8}^{+0.8}$ & $0.07_{-0.04}^{+0.04}$ \\  [6pt]
		iso\_{11.0}$^b$ & $0.11_{-0.02}^{+0.02}$ & $0.045_{-0.001}^{+0.001}$ & $2.4_{-0.4}^{+0.4}$ & $0.07_{-0.02}^{+0.02}$ \\
		\hline
	\end{tabular}
	\begin{tablenotes}[normal]
	 \item \textit{Notes.} $^a$Isolated MGs ($\log(M_{*}/M_{\sun})\geq11.3$). $^b$Isolated MGs ($\log(M_{*}/M_{\sun})\geq11.0$). In the calculation, we exclude galaxies with $\log(M_{*}/M_{\odot})\geq 11$ to avoid possible selection biases.
    \end{tablenotes}
\end{table}

\section{Summary and Conclusions}
We have searched for proto-cluster cores at $z\sim 2$ in $\sim 1.5\, \mathrm{deg}^{2}$ of the COSMOS field by using pairs of MGs ($\log(M_{*}/M_{\sun})\geq11$) as tracers, and examined properties of member galaxies in the cores. The main results are as follows.
\begin{enumerate}
  \item We find 75 pairs of MGs whose separations are $<30\arcsec$, among which 54\% are estimated to be real.
  
  \item A clustering analysis finds that the average mass of DMHs hosting the pairs is $2.6_{-0.8}^{+0.9}\times 10^{13}\, M_\mathrm{\sun}$, and $4.0^{+1.8}_{-1.5}\times 10^{13}\, M_\mathrm{\sun}$ after contamination correction. Using the extended Press-Schechter model, we also calculate the descendant DMH mass and confirm that the pairs are typically progenitors of Virgo or Fornax-like clusters. 
  
  \item The IllustrisTNG simulation shows pairs of MGs are good tracers of DMHs which are massive enough to be regraded as proto-cluster cores. At a fixed stellar mass, the median mass of DMHs which host pairs is larger by 0.15 to 0.3 dex than those of DMHs which do not. We also find that more than 50\% of DMHs with $2.6\times 10^{13}\, M_{\sun}$ host pairs, which is consistent with the completeness estimated from the halo mass function. Since the pair-host fraction is a monotonically increasing function of $M_\mathrm{DMH}$, the most massive DMHs can be traced by pairs at $z=2$. We trace merger trees from $z=2$ to $z=0$ to identify descendants of pair-host haloes. We find that 100 independent DMHs which host pairs at $z=2$ become 89 independent DMHs at $z=0$. At $z=0$, the numbers of descendants of pair-host haloes (and all haloes in the simulation box) classified as Fornax-like, Virgo-like and Coma-like clusters are 38 (235), 22 (42), 3 (3), respectively, resulting in $16\%$, $52\%$ and $100\%$ completeness for each type. This suggests that a pair of MGs can trace progenitors of both the most massive clusters and less massive ones.
  
  \item The member galaxies of the cores have a more top-heavy SMF than the field except for quiescent galaxies. When normalised by total mass, the ratio of SMFs between cores and the field is below unity at $\log(M_{*}/M_{\sun})\lesssim 10$ and above unity at $\log(M_{*}/M_{\sun})\gtrsim 10$. The low ratio at $\log(M_{*}/M_{\sun})\lesssim 10$, if real, may indicate that low-mass galaxies in cores are more likely to be prevented from forming stars, or destroyed by mergers and/or tidal disruption than field galaxies. On the other hand, the star formation of high-mass galaxies may be enhanced by the early formation of massive DMHs and/or more frequent mergers. These trends are similar to SMFs in previous studies focusing on known (proto-)clusters and local high-density regions.
  
  \item The quiescent fraction of the member galaxies in the cores is higher than that of the field at $\log(M_{*}/M_{\sun})\lesssim 10.6$. The quiescent fraction averaged over the whole mass range $9<\log(M_{*}/M_{\sun})<11$ is $0.17_{-0.04}^{+0.04}$, which is three time higher than that of the field. We also calculate the environmental quenching efficiency (QE) and find that the QE in the cores is comparable to that of mature clusters at $z\sim 1.6$ in the literature. This supports a scenario that cluster environments have not quenched galaxies significantly until $z\sim 1.5$ when a whole proto-cluster region starts to collapse, although excess quenching is already seen in cores.
\end{enumerate}

We have statistically shown that proto-cluster cores at $z\sim 2$ have similar properties to mature clusters at $z\lesssim 1.5$ in terms of an excess of high-mass galaxies and a higher fraction of low-mass quiescent galaxies. These results suggest that stellar mass assembly and quenching are accelerated as early as $z\sim 2$ in proto-cluster cores. To investigate other properties further, spectroscopic confirmation of the individual cores is needed. Our core sample presents good targets for spectroscopic surveys like the Subaru Prime Focus Spectrograph survey \citep{Takada2014}. If we derive precise redshifts of member galaxies, we can calculate individual DMH masses from their velocity dispersions. We can also reveal detailed star-forming activities with spectroscopic data, and thus the formation history of cluster galaxies \citep{Harikane2019}.

The method presented in this paper can be applied to other survey data with stellar mass and photo-\textit{z} estimates. Therefore, combining wide field surveys like the Subaru Hyper Suprime-Cam survey (HSC-SSP), we can construct a much larger core sample over a wide redshift range.

\section*{Acknowledgements}
We would like to thank Drs. Haruka Kusakabe, Taku Okamura, Mr. Takahiro Sudoh and Ms. Hinako Goto for helpful comments and discussions. We also would like to thank the anonymous referee for very constructive comments. Based on data products from observations made with ESO Telescopes at the La Silla Paranal Observatory under ESO programme ID 179.A-2005 and on data products produced by TERAPIX and the Cambridge Astronomy Survey Unit on behalf of the UltraVISTA consortium. We acknowledge the team of the IllustrisTNG project (\url{https://www.tng-project.org/}). We use the following open source software packages for our analysis: \texttt{numpy} \citep{numpy:2011}, \texttt{pandas} \citep{pandas:2010}, \texttt{scipy} \citep{scipy:2001}, \texttt{astropy} \citep{astropy:2013,astropy:2018} and \texttt{matplotlib} \citep{matplotlib:2007}. RM acknowledges a Japan Society for the Promotion of Science (JSPS) Fellowship at Japan. This work is supported in part by JSPS KAKENHI Grant Numbers JP19K03924 (KS) and JP18J40088 (RM).




\bibliographystyle{mnras}
\bibliography{pCLcore} 

\begin{thebibliography}{}
\makeatletter
\relax
\def\mn@urlcharsother{\let\do\@makeother \do\$\do\&\do\#\do\^\do\_\do\%\do\~}
\def\mn@doi{\begingroup\mn@urlcharsother \@ifnextchar [ {\mn@doi@}
  {\mn@doi@[]}}
\def\mn@doi@[#1]#2{\def\@tempa{#1}\ifx\@tempa\@empty \href
  {http://dx.doi.org/#2} {doi:#2}\else \href {http://dx.doi.org/#2} {#1}\fi
  \endgroup}
\def\mn@eprint#1#2{\mn@eprint@#1:#2::\@nil}
\def\mn@eprint@arXiv#1{\href {http://arxiv.org/abs/#1} {{\tt arXiv:#1}}}
\def\mn@eprint@dblp#1{\href {http://dblp.uni-trier.de/rec/bibtex/#1.xml}
  {dblp:#1}}
\def\mn@eprint@#1:#2:#3:#4\@nil{\def\@tempa {#1}\def\@tempb {#2}\def\@tempc
  {#3}\ifx \@tempc \@empty \let \@tempc \@tempb \let \@tempb \@tempa \fi \ifx
  \@tempb \@empty \def\@tempb {arXiv}\fi \@ifundefined
  {mn@eprint@\@tempb}{\@tempb:\@tempc}{\expandafter \expandafter \csname
  mn@eprint@\@tempb\endcsname \expandafter{\@tempc}}}

\bibitem[\protect\citeauthoryear{{Arnouts} et~al.,}{{Arnouts}
  et~al.}{2002}]{Arnouts2002}
{Arnouts} S.,  et~al., 2002, \mn@doi [\mnras]
  {10.1046/j.1365-8711.2002.04988.x}, \href
  {https://ui.adsabs.harvard.edu/abs/2002MNRAS.329..355A} {329, 355}

\bibitem[\protect\citeauthoryear{{Astropy Collaboration} et~al.,}{{Astropy
  Collaboration} et~al.}{2013}]{astropy:2013}
{Astropy Collaboration} et~al., 2013, \mn@doi [\aap]
  {10.1051/0004-6361/201322068}, \href
  {https://ui.adsabs.harvard.edu/abs/2013A&A...558A..33A} {558, A33}

\bibitem[\protect\citeauthoryear{{Astropy Collaboration} et~al.,}{{Astropy
  Collaboration} et~al.}{2018}]{astropy:2018}
{Astropy Collaboration} et~al., 2018, \mn@doi [\aj] {10.3847/1538-3881/aabc4f},
  \href {https://ui.adsabs.harvard.edu/abs/2018AJ....156..123A} {156, 123}

\bibitem[\protect\citeauthoryear{{Balogh} et~al.,}{{Balogh}
  et~al.}{2014}]{Balogh2014}
{Balogh} M.~L.,  et~al., 2014, \mn@doi [\mnras] {10.1093/mnras/stu1332}, \href
  {https://ui.adsabs.harvard.edu/abs/2014MNRAS.443.2679B} {443, 2679}

\bibitem[\protect\citeauthoryear{{Behroozi}, {Wechsler}  \&
  {Conroy}}{{Behroozi} et~al.}{2013}]{Behroozi2013}
{Behroozi} P.~S.,  {Wechsler} R.~H.,   {Conroy} C.,  2013, \mn@doi [\apj]
  {10.1088/0004-637X/770/1/57}, \href
  {https://ui.adsabs.harvard.edu/abs/2013ApJ...770...57B} {770, 57}

\bibitem[\protect\citeauthoryear{{B{\'e}thermin} et~al.,}{{B{\'e}thermin}
  et~al.}{2014}]{Bethermin2014}
{B{\'e}thermin} M.,  et~al., 2014, \mn@doi [\aap]
  {10.1051/0004-6361/201423451}, \href
  {https://ui.adsabs.harvard.edu/abs/2014A&A...567A.103B} {567, A103}

\bibitem[\protect\citeauthoryear{{Bower}, {Kodama}  \& {Terlevich}}{{Bower}
  et~al.}{1998}]{Bower1998}
{Bower} R.~G.,  {Kodama} T.,   {Terlevich} A.,  1998, \mn@doi [\mnras]
  {10.1046/j.1365-8711.1998.01868.x}, \href
  {https://ui.adsabs.harvard.edu/abs/1998MNRAS.299.1193B} {299, 1193}

\bibitem[\protect\citeauthoryear{{Chabrier}}{{Chabrier}}{2003}]{Chabrier2003}
{Chabrier} G.,  2003, \mn@doi [\pasp] {10.1086/376392}, \href
  {https://ui.adsabs.harvard.edu/abs/2003PASP..115..763C} {115, 763}

\bibitem[\protect\citeauthoryear{{Chartab} et~al.,}{{Chartab}
  et~al.}{2020}]{Chartab2020}
{Chartab} N.,  et~al., 2020, \mn@doi [\apj] {10.3847/1538-4357/ab61fd}, \href
  {https://ui.adsabs.harvard.edu/abs/2020ApJ...890....7C} {890, 7}

\bibitem[\protect\citeauthoryear{{Chiang}, {Overzier}  \& {Gebhardt}}{{Chiang}
  et~al.}{2013}]{Chiang2013}
{Chiang} Y.-K.,  {Overzier} R.,   {Gebhardt} K.,  2013, \mn@doi [\apj]
  {10.1088/0004-637X/779/2/127}, \href
  {https://ui.adsabs.harvard.edu/abs/2013ApJ...779..127C} {779, 127}

\bibitem[\protect\citeauthoryear{{Chiang}, {Overzier}  \& {Gebhardt}}{{Chiang}
  et~al.}{2014}]{Chiang2014}
{Chiang} Y.-K.,  {Overzier} R.,   {Gebhardt} K.,  2014, \mn@doi [\apjl]
  {10.1088/2041-8205/782/1/L3}, \href
  {https://ui.adsabs.harvard.edu/abs/2014ApJ...782L...3C} {782, L3}

\bibitem[\protect\citeauthoryear{{Chiang} et~al.,}{{Chiang}
  et~al.}{2015}]{Chiang2015}
{Chiang} Y.-K.,  et~al., 2015, \mn@doi [\apj] {10.1088/0004-637X/808/1/37},
  \href {https://ui.adsabs.harvard.edu/abs/2015ApJ...808...37C} {808, 37}

\bibitem[\protect\citeauthoryear{{Chiang}, {Overzier}, {Gebhardt}  \&
  {Henriques}}{{Chiang} et~al.}{2017}]{Chiang2017}
{Chiang} Y.-K.,  {Overzier} R.~A.,  {Gebhardt} K.,   {Henriques} B.,  2017,
  \mn@doi [\apjl] {10.3847/2041-8213/aa7e7b}, \href
  {https://ui.adsabs.harvard.edu/abs/2017ApJ...844L..23C} {844, L23}

\bibitem[\protect\citeauthoryear{{Contini}, {Gu}, {Ge}, {Rhee}, {Yi}  \&
  {Kang}}{{Contini} et~al.}{2020}]{Contini2020}
{Contini} E.,  {Gu} Q.,  {Ge} X.,  {Rhee} J.,  {Yi} S.~K.,   {Kang} X.,  2020,
  \mn@doi [\apj] {10.3847/1538-4357/ab6730}, \href
  {https://ui.adsabs.harvard.edu/abs/2020ApJ...889..156C} {889, 156}

\bibitem[\protect\citeauthoryear{{Cooke}, {Hatch}, {Muldrew}, {Rigby}  \&
  {Kurk}}{{Cooke} et~al.}{2014}]{Cooke2014}
{Cooke} E.~A.,  {Hatch} N.~A.,  {Muldrew} S.~I.,  {Rigby} E.~E.,   {Kurk}
  J.~D.,  2014, \mn@doi [\mnras] {10.1093/mnras/stu522}, \href
  {https://ui.adsabs.harvard.edu/abs/2014MNRAS.440.3262C} {440, 3262}

\bibitem[\protect\citeauthoryear{{Cooke} et~al.,}{{Cooke}
  et~al.}{2016}]{Cooke2016}
{Cooke} E.~A.,  et~al., 2016, \mn@doi [\apj] {10.3847/0004-637X/816/2/83},
  \href {https://ui.adsabs.harvard.edu/abs/2016ApJ...816...83C} {816, 83}

\bibitem[\protect\citeauthoryear{{Cowan} \& {Ivezi{\'c}}}{{Cowan} \&
  {Ivezi{\'c}}}{2008}]{Cowan2008}
{Cowan} N.~B.,  {Ivezi{\'c}} {\v{Z}}.,  2008, \mn@doi [\apjl] {10.1086/528986},
  \href {https://ui.adsabs.harvard.edu/abs/2008ApJ...674L..13C} {674, L13}

\bibitem[\protect\citeauthoryear{{Croom} \& {Shanks}}{{Croom} \&
  {Shanks}}{1999}]{Croom1999}
{Croom} S.~M.,  {Shanks} T.,  1999, \mn@doi [\mnras]
  {10.1046/j.1365-8711.1999.02232.x}, \href
  {https://ui.adsabs.harvard.edu/abs/1999MNRAS.303..411C} {303, 411}

\bibitem[\protect\citeauthoryear{{Darvish}, {Mobasher}, {Sobral}, {Scoville}
  \& {Aragon-Calvo}}{{Darvish} et~al.}{2015}]{Darvish2015}
{Darvish} B.,  {Mobasher} B.,  {Sobral} D.,  {Scoville} N.,   {Aragon-Calvo}
  M.,  2015, \mn@doi [\apj] {10.1088/0004-637X/805/2/121}, \href
  {https://ui.adsabs.harvard.edu/abs/2015ApJ...805..121D} {805, 121}

\bibitem[\protect\citeauthoryear{{Diemer}}{{Diemer}}{2018}]{Diemer2018}
{Diemer} B.,  2018, \mn@doi [\apjs] {10.3847/1538-4365/aaee8c}, \href
  {https://ui.adsabs.harvard.edu/abs/2018ApJS..239...35D} {239, 35}

\bibitem[\protect\citeauthoryear{{Diener} et~al.,}{{Diener}
  et~al.}{2013}]{Diener2013}
{Diener} C.,  et~al., 2013, \mn@doi [\apj] {10.1088/0004-637X/765/2/109}, \href
  {https://ui.adsabs.harvard.edu/abs/2013ApJ...765..109D} {765, 109}

\bibitem[\protect\citeauthoryear{{Diener} et~al.,}{{Diener}
  et~al.}{2015}]{Diener2015}
{Diener} C.,  et~al., 2015, \mn@doi [\apj] {10.1088/0004-637X/802/1/31}, \href
  {https://ui.adsabs.harvard.edu/abs/2015ApJ...802...31D} {802, 31}

\bibitem[\protect\citeauthoryear{{Dressler}}{{Dressler}}{1980}]{Dressler1980}
{Dressler} A.,  1980, \mn@doi [\apj] {10.1086/157753}, \href
  {https://ui.adsabs.harvard.edu/abs/1980ApJ...236..351D} {236, 351}

\bibitem[\protect\citeauthoryear{{Efstathiou}, {Bernstein}, {Katz}, {Tyson}  \&
  {Guhathakurta}}{{Efstathiou} et~al.}{1991}]{Efstathiou1991}
{Efstathiou} G.,  {Bernstein} G.,  {Katz} N.,  {Tyson} J.~A.,   {Guhathakurta}
  P.,  1991, \mn@doi [\apjl] {10.1086/186170}, \href
  {https://ui.adsabs.harvard.edu/abs/1991ApJ...380L..47E} {380, L47}

\bibitem[\protect\citeauthoryear{{Eisenstein} \& {Hu}}{{Eisenstein} \&
  {Hu}}{1999}]{Eisenstein1999}
{Eisenstein} D.~J.,  {Hu} W.,  1999, \mn@doi [\apj] {10.1086/306640}, \href
  {https://ui.adsabs.harvard.edu/abs/1999ApJ...511....5E} {511, 5}

\bibitem[\protect\citeauthoryear{{Goto}, {Yamauchi}, {Fujita}, {Okamura},
  {Sekiguchi}, {Smail}, {Bernardi}  \& {Gomez}}{{Goto} et~al.}{2003}]{Goto2003}
{Goto} T.,  {Yamauchi} C.,  {Fujita} Y.,  {Okamura} S.,  {Sekiguchi} M.,
  {Smail} I.,  {Bernardi} M.,   {Gomez} P.~L.,  2003, \mn@doi [\mnras]
  {10.1046/j.1365-2966.2003.07114.x}, \href
  {https://ui.adsabs.harvard.edu/abs/2003MNRAS.346..601G} {346, 601}

\bibitem[\protect\citeauthoryear{{Groth} \& {Peebles}}{{Groth} \&
  {Peebles}}{1977}]{Groth1977}
{Groth} E.~J.,  {Peebles} P.~J.~E.,  1977, \mn@doi [\apj] {10.1086/155588},
  \href {https://ui.adsabs.harvard.edu/abs/1977ApJ...217..385G} {217, 385}

\bibitem[\protect\citeauthoryear{{Gunn} \& {Gott}}{{Gunn} \&
  {Gott}}{1972}]{Gunn1972}
{Gunn} J.~E.,  {Gott} J.~Richard I.,  1972, \mn@doi [\apj] {10.1086/151605},
  \href {https://ui.adsabs.harvard.edu/abs/1972ApJ...176....1G} {176, 1}

\bibitem[\protect\citeauthoryear{{Harikane} et~al.,}{{Harikane}
  et~al.}{2019}]{Harikane2019}
{Harikane} Y.,  et~al., 2019, \mn@doi [\apj] {10.3847/1538-4357/ab2cd5}, \href
  {https://ui.adsabs.harvard.edu/abs/2019ApJ...883..142H} {883, 142}

\bibitem[\protect\citeauthoryear{{Hatch} et~al.,}{{Hatch}
  et~al.}{2011}]{Hatch2011}
{Hatch} N.~A.,  et~al., 2011, \mn@doi [\mnras]
  {10.1111/j.1365-2966.2010.17538.x}, \href
  {https://ui.adsabs.harvard.edu/abs/2011MNRAS.410.1537H} {410, 1537}

\bibitem[\protect\citeauthoryear{{Hatch} et~al.,}{{Hatch}
  et~al.}{2014}]{Hatch2014}
{Hatch} N.~A.,  et~al., 2014, \mn@doi [\mnras] {10.1093/mnras/stu1725}, \href
  {https://ui.adsabs.harvard.edu/abs/2014MNRAS.445..280H} {445, 280}

\bibitem[\protect\citeauthoryear{{Hunter}}{{Hunter}}{2007}]{matplotlib:2007}
{Hunter} J.~D.,  2007, \mn@doi [Computing in Science and Engineering]
  {10.1109/MCSE.2007.55}, \href
  {https://ui.adsabs.harvard.edu/abs/2007CSE.....9...90H} {9, 90}

\bibitem[\protect\citeauthoryear{{Ilbert} et~al.,}{{Ilbert}
  et~al.}{2006}]{Ilbert2006}
{Ilbert} O.,  et~al., 2006, \mn@doi [\aap] {10.1051/0004-6361:20065138}, \href
  {https://ui.adsabs.harvard.edu/abs/2006A&A...457..841I} {457, 841}

\bibitem[\protect\citeauthoryear{{Ji}, {Giavalisco}, {Williams}, {Faber},
  {Ferguson}, {Guo}, {Liu}  \& {Lee}}{{Ji} et~al.}{2018}]{Ji2018}
{Ji} Z.,  {Giavalisco} M.,  {Williams} C.~C.,  {Faber} S.~M.,  {Ferguson}
  H.~C.,  {Guo} Y.,  {Liu} T.,   {Lee} B.,  2018, \mn@doi [\apj]
  {10.3847/1538-4357/aacc2c}, \href
  {https://ui.adsabs.harvard.edu/abs/2018ApJ...862..135J} {862, 135}

\bibitem[\protect\citeauthoryear{Jones, Oliphant, Peterson  et~al.}{Jones
  et~al.}{2001}]{scipy:2001}
Jones E.,  Oliphant T.,  Peterson P.,   et~al., 2001, {SciPy}: Open source
  scientific tools for {Python}, \url {http://www.scipy.org/}

\bibitem[\protect\citeauthoryear{{Kawinwanichakij} et~al.,}{{Kawinwanichakij}
  et~al.}{2016}]{Kawinwanichakij2016}
{Kawinwanichakij} L.,  et~al., 2016, \mn@doi [\apj]
  {10.3847/0004-637X/817/1/9}, \href
  {https://ui.adsabs.harvard.edu/abs/2016ApJ...817....9K} {817, 9}

\bibitem[\protect\citeauthoryear{{Kawinwanichakij} et~al.,}{{Kawinwanichakij}
  et~al.}{2017}]{Kawinwanichakij2017}
{Kawinwanichakij} L.,  et~al., 2017, \mn@doi [\apj] {10.3847/1538-4357/aa8b75},
  \href {https://ui.adsabs.harvard.edu/abs/2017ApJ...847..134K} {847, 134}

\bibitem[\protect\citeauthoryear{{Kravtsov} \& {Borgani}}{{Kravtsov} \&
  {Borgani}}{2012}]{Kravtsov_Borgani_2012}
{Kravtsov} A.~V.,  {Borgani} S.,  2012, \mn@doi [\araa]
  {10.1146/annurev-astro-081811-125502}, \href
  {https://ui.adsabs.harvard.edu/abs/2012ARA&A..50..353K} {50, 353}

\bibitem[\protect\citeauthoryear{{Laigle} et~al.,}{{Laigle}
  et~al.}{2016}]{Laigle2016}
{Laigle} C.,  et~al., 2016, \mn@doi [\apjs] {10.3847/0067-0049/224/2/24}, \href
  {https://ui.adsabs.harvard.edu/abs/2016ApJS..224...24L} {224, 24}

\bibitem[\protect\citeauthoryear{{Landy} \& {Szalay}}{{Landy} \&
  {Szalay}}{1993}]{Landy1993}
{Landy} S.~D.,  {Szalay} A.~S.,  1993, \mn@doi [\apj] {10.1086/172900}, \href
  {https://ui.adsabs.harvard.edu/abs/1993ApJ...412...64L} {412, 64}

\bibitem[\protect\citeauthoryear{{Lee-Brown} et~al.,}{{Lee-Brown}
  et~al.}{2017}]{Lee-Brown2017}
{Lee-Brown} D.~B.,  et~al., 2017, \mn@doi [\apj] {10.3847/1538-4357/aa7948},
  \href {https://ui.adsabs.harvard.edu/abs/2017ApJ...844...43L} {844, 43}

\bibitem[\protect\citeauthoryear{{Lovell}, {Thomas}  \& {Wilkins}}{{Lovell}
  et~al.}{2018}]{Lovell2018}
{Lovell} C.~C.,  {Thomas} P.~A.,   {Wilkins} S.~M.,  2018, \mn@doi [\mnras]
  {10.1093/mnras/stx3090}, \href
  {https://ui.adsabs.harvard.edu/abs/2018MNRAS.474.4612L} {474, 4612}

\bibitem[\protect\citeauthoryear{{Malavasi}, {Pozzetti}, {Cucciati}, {Bardelli}
   \& {Cimatti}}{{Malavasi} et~al.}{2016}]{Malavasi2016}
{Malavasi} N.,  {Pozzetti} L.,  {Cucciati} O.,  {Bardelli} S.,   {Cimatti} A.,
  2016, \mn@doi [\aap] {10.1051/0004-6361/201526718}, \href
  {https://ui.adsabs.harvard.edu/abs/2016A&A...585A.116M} {585, A116}

\bibitem[\protect\citeauthoryear{{Marinacci} et~al.,}{{Marinacci}
  et~al.}{2018}]{Marinacci2018}
{Marinacci} F.,  et~al., 2018, \mn@doi [\mnras] {10.1093/mnras/sty2206}, \href
  {https://ui.adsabs.harvard.edu/abs/2018MNRAS.480.5113M} {480, 5113}

\bibitem[\protect\citeauthoryear{{Martel}, {Barai}  \& {Brito}}{{Martel}
  et~al.}{2012}]{Martel2012}
{Martel} H.,  {Barai} P.,   {Brito} W.,  2012, \mn@doi [\apj]
  {10.1088/0004-637X/757/1/48}, \href
  {https://ui.adsabs.harvard.edu/abs/2012ApJ...757...48M} {757, 48}

\bibitem[\protect\citeauthoryear{{Martini}}{{Martini}}{2004}]{Martini2004}
{Martini} P.,  2004, in {Ho} L.~C.,  ed., Coevolution of Black Holes and
  Galaxies. p.~169 (\mn@eprint {arXiv} {astro-ph/0304009})

\bibitem[\protect\citeauthoryear{McKinney}{McKinney}{2010}]{pandas:2010}
McKinney W.,  2010, in van~der Walt S.,  Millman J.,  eds, Proceedings of the
  9th Python in Science Conference. pp 51 -- 56

\bibitem[\protect\citeauthoryear{{Miller} et~al.,}{{Miller}
  et~al.}{2018}]{Miller2018}
{Miller} T.~B.,  et~al., 2018, \mn@doi [\nat] {10.1038/s41586-018-0025-2},
  \href {https://ui.adsabs.harvard.edu/abs/2018Natur.556..469M} {556, 469}

\bibitem[\protect\citeauthoryear{{Moore}, {Lake}  \& {Katz}}{{Moore}
  et~al.}{1998}]{Moore1998}
{Moore} B.,  {Lake} G.,   {Katz} N.,  1998, \mn@doi [\apj] {10.1086/305264},
  \href {https://ui.adsabs.harvard.edu/abs/1998ApJ...495..139M} {495, 139}

\bibitem[\protect\citeauthoryear{{Muldrew}, {Hatch}  \& {Cooke}}{{Muldrew}
  et~al.}{2015}]{Muldrew2015}
{Muldrew} S.~I.,  {Hatch} N.~A.,   {Cooke} E.~A.,  2015, \mn@doi [\mnras]
  {10.1093/mnras/stv1449}, \href
  {https://ui.adsabs.harvard.edu/abs/2015MNRAS.452.2528M} {452, 2528}

\bibitem[\protect\citeauthoryear{{Muldrew}, {Hatch}  \& {Cooke}}{{Muldrew}
  et~al.}{2018}]{Muldrew2018}
{Muldrew} S.~I.,  {Hatch} N.~A.,   {Cooke} E.~A.,  2018, \mn@doi [\mnras]
  {10.1093/mnras/stx2454}, \href
  {https://ui.adsabs.harvard.edu/abs/2018MNRAS.473.2335M} {473, 2335}

\bibitem[\protect\citeauthoryear{{Naiman} et~al.,}{{Naiman}
  et~al.}{2018}]{Naiman2018}
{Naiman} J.~P.,  et~al., 2018, \mn@doi [\mnras] {10.1093/mnras/sty618}, \href
  {https://ui.adsabs.harvard.edu/abs/2018MNRAS.477.1206N} {477, 1206}

\bibitem[\protect\citeauthoryear{{Nantais} et~al.,}{{Nantais}
  et~al.}{2016}]{Nantais2016}
{Nantais} J.~B.,  et~al., 2016, \mn@doi [\aap] {10.1051/0004-6361/201628663},
  \href {https://ui.adsabs.harvard.edu/abs/2016A&A...592A.161N} {592, A161}

\bibitem[\protect\citeauthoryear{{Nantais} et~al.,}{{Nantais}
  et~al.}{2017}]{Nantais2017}
{Nantais} J.~B.,  et~al., 2017, \mn@doi [\mnras] {10.1093/mnrasl/slw224}, \href
  {https://ui.adsabs.harvard.edu/abs/2017MNRAS.465L.104N} {465, L104}

\bibitem[\protect\citeauthoryear{{Nelson} et~al.,}{{Nelson}
  et~al.}{2018}]{Nelson2018}
{Nelson} D.,  et~al., 2018, \mn@doi [\mnras] {10.1093/mnras/stx3040}, \href
  {https://ui.adsabs.harvard.edu/abs/2018MNRAS.475..624N} {475, 624}

\bibitem[\protect\citeauthoryear{{Newman}, {Ellis}, {Andreon}, {Treu},
  {Raichoor}  \& {Trinchieri}}{{Newman} et~al.}{2014}]{Newman2014}
{Newman} A.~B.,  {Ellis} R.~S.,  {Andreon} S.,  {Treu} T.,  {Raichoor} A.,
  {Trinchieri} G.,  2014, \mn@doi [\apj] {10.1088/0004-637X/788/1/51}, \href
  {https://ui.adsabs.harvard.edu/abs/2014ApJ...788...51N} {788, 51}

\bibitem[\protect\citeauthoryear{{Oteo} et~al.,}{{Oteo}
  et~al.}{2018}]{Oteo2018}
{Oteo} I.,  et~al., 2018, \mn@doi [\apj] {10.3847/1538-4357/aaa1f1}, \href
  {https://ui.adsabs.harvard.edu/abs/2018ApJ...856...72O} {856, 72}

\bibitem[\protect\citeauthoryear{{Ouchi} et~al.,}{{Ouchi}
  et~al.}{2003}]{Ouchi2003}
{Ouchi} M.,  et~al., 2003, \mn@doi [\apj] {10.1086/344476}, \href
  {https://ui.adsabs.harvard.edu/abs/2003ApJ...582...60O} {582, 60}

\bibitem[\protect\citeauthoryear{{Overzier}}{{Overzier}}{2016}]{Overzier2016}
{Overzier} R.~A.,  2016, \mn@doi [\aapr] {10.1007/s00159-016-0100-3}, \href
  {https://ui.adsabs.harvard.edu/abs/2016A&ARv..24...14O} {24, 14}

\bibitem[\protect\citeauthoryear{{Papovich} et~al.,}{{Papovich}
  et~al.}{2018}]{Papovich2018}
{Papovich} C.,  et~al., 2018, \mn@doi [\apj] {10.3847/1538-4357/aaa766}, \href
  {https://ui.adsabs.harvard.edu/abs/2018ApJ...854...30P} {854, 30}

\bibitem[\protect\citeauthoryear{{Peebles}}{{Peebles}}{1975}]{Peebles1975}
{Peebles} P.~J.~E.,  1975, \mn@doi [\apj] {10.1086/153450}, \href
  {https://ui.adsabs.harvard.edu/abs/1975ApJ...196..647P} {196, 647}

\bibitem[\protect\citeauthoryear{{Peebles}}{{Peebles}}{1980}]{Peebles1980}
{Peebles} P.~J.~E.,  1980, {The large-scale structure of the universe}

\bibitem[\protect\citeauthoryear{{Peng} et~al.,}{{Peng}
  et~al.}{2010}]{Peng2010}
{Peng} Y.-j.,  et~al., 2010, \mn@doi [\apj] {10.1088/0004-637X/721/1/193},
  \href {https://ui.adsabs.harvard.edu/abs/2010ApJ...721..193P} {721, 193}

\bibitem[\protect\citeauthoryear{{Pillepich} et~al.,}{{Pillepich}
  et~al.}{2018a}]{Pillepich2018a}
{Pillepich} A.,  et~al., 2018a, \mn@doi [\mnras] {10.1093/mnras/stx2656}, \href
  {https://ui.adsabs.harvard.edu/abs/2018MNRAS.473.4077P} {473, 4077}

\bibitem[\protect\citeauthoryear{{Pillepich} et~al.,}{{Pillepich}
  et~al.}{2018b}]{Pillepich2018c}
{Pillepich} A.,  et~al., 2018b, \mn@doi [\mnras] {10.1093/mnras/stx3112}, \href
  {https://ui.adsabs.harvard.edu/abs/2018MNRAS.475..648P} {475, 648}

\bibitem[\protect\citeauthoryear{{Quadri}, {Williams}, {Franx}  \&
  {Hildebrandt}}{{Quadri} et~al.}{2012}]{Quadri2012}
{Quadri} R.~F.,  {Williams} R.~J.,  {Franx} M.,   {Hildebrandt} H.,  2012,
  \mn@doi [\apj] {10.1088/0004-637X/744/2/88}, \href
  {https://ui.adsabs.harvard.edu/abs/2012ApJ...744...88Q} {744, 88}

\bibitem[\protect\citeauthoryear{{Roche} \& {Eales}}{{Roche} \&
  {Eales}}{1999}]{Roche1999}
{Roche} N.,  {Eales} S.~A.,  1999, \mn@doi [\mnras]
  {10.1046/j.1365-8711.1999.02652.x}, \href
  {https://ui.adsabs.harvard.edu/abs/1999MNRAS.307..703R} {307, 703}

\bibitem[\protect\citeauthoryear{{Rodr{\'\i}guez-Mu{\~n}oz}
  et~al.,}{{Rodr{\'\i}guez-Mu{\~n}oz} et~al.}{2019}]{Rodriguez2019}
{Rodr{\'\i}guez-Mu{\~n}oz} L.,  et~al., 2019, \mn@doi [\mnras]
  {10.1093/mnras/sty3335}, \href
  {https://ui.adsabs.harvard.edu/abs/2019MNRAS.485..586R} {485, 586}

\bibitem[\protect\citeauthoryear{{Scoville} et~al.,}{{Scoville}
  et~al.}{2007}]{Scoville2007}
{Scoville} N.,  et~al., 2007, \mn@doi [\apjs] {10.1086/516580}, \href
  {https://ui.adsabs.harvard.edu/abs/2007ApJS..172...38S} {172, 38}

\bibitem[\protect\citeauthoryear{{Sheth} \& {Tormen}}{{Sheth} \&
  {Tormen}}{1999}]{Sheth1999}
{Sheth} R.~K.,  {Tormen} G.,  1999, \mn@doi [\mnras]
  {10.1046/j.1365-8711.1999.02692.x}, \href
  {https://ui.adsabs.harvard.edu/abs/1999MNRAS.308..119S} {308, 119}

\bibitem[\protect\citeauthoryear{{Shimakawa} et~al.,}{{Shimakawa}
  et~al.}{2018}]{Shimakawa2018}
{Shimakawa} R.,  et~al., 2018, \mn@doi [\mnras] {10.1093/mnras/stx2494}, \href
  {https://ui.adsabs.harvard.edu/abs/2018MNRAS.473.1977S} {473, 1977}

\bibitem[\protect\citeauthoryear{{Springel} et~al.,}{{Springel}
  et~al.}{2018}]{Springel2018}
{Springel} V.,  et~al., 2018, \mn@doi [\mnras] {10.1093/mnras/stx3304}, \href
  {https://ui.adsabs.harvard.edu/abs/2018MNRAS.475..676S} {475, 676}

\bibitem[\protect\citeauthoryear{{Takada} et~al.,}{{Takada}
  et~al.}{2014}]{Takada2014}
{Takada} M.,  et~al., 2014, \mn@doi [\pasj] {10.1093/pasj/pst019}, \href
  {https://ui.adsabs.harvard.edu/abs/2014PASJ...66R...1T} {66, R1}

\bibitem[\protect\citeauthoryear{{Tinker}, {Robertson}, {Kravtsov}, {Klypin},
  {Warren}, {Yepes}  \& {Gottl{\"o}ber}}{{Tinker} et~al.}{2010}]{Tinker2010}
{Tinker} J.~L.,  {Robertson} B.~E.,  {Kravtsov} A.~V.,  {Klypin} A.,  {Warren}
  M.~S.,  {Yepes} G.,   {Gottl{\"o}ber} S.,  2010, \mn@doi [\apj]
  {10.1088/0004-637X/724/2/878}, \href
  {https://ui.adsabs.harvard.edu/abs/2010ApJ...724..878T} {724, 878}

\bibitem[\protect\citeauthoryear{{Tomczak} et~al.,}{{Tomczak}
  et~al.}{2017}]{Tomczak2017}
{Tomczak} A.~R.,  et~al., 2017, \mn@doi [\mnras] {10.1093/mnras/stx2245}, \href
  {https://ui.adsabs.harvard.edu/abs/2017MNRAS.472.3512T} {472, 3512}

\bibitem[\protect\citeauthoryear{{Toshikawa} et~al.,}{{Toshikawa}
  et~al.}{2018}]{Toshikawa2018}
{Toshikawa} J.,  et~al., 2018, \mn@doi [\pasj] {10.1093/pasj/psx102}, \href
  {https://ui.adsabs.harvard.edu/abs/2018PASJ...70S..12T} {70, S12}

\bibitem[\protect\citeauthoryear{{Uchiyama} et~al.,}{{Uchiyama}
  et~al.}{2018}]{Uchiyama2018}
{Uchiyama} H.,  et~al., 2018, \mn@doi [\pasj] {10.1093/pasj/psx112}, \href
  {https://ui.adsabs.harvard.edu/abs/2018PASJ...70S..32U} {70, S32}

\bibitem[\protect\citeauthoryear{{Uchiyama} et~al.,}{{Uchiyama}
  et~al.}{2019}]{Uchiyama2019}
{Uchiyama} H.,  et~al., 2019, \mn@doi [\apj] {10.3847/1538-4357/aaef7b}, \href
  {https://ui.adsabs.harvard.edu/abs/2019ApJ...870...45U} {870, 45}

\bibitem[\protect\citeauthoryear{{Venemans} et~al.,}{{Venemans}
  et~al.}{2007}]{Venemans2007}
{Venemans} B.~P.,  et~al., 2007, \mn@doi [\aap] {10.1051/0004-6361:20053941},
  \href {https://ui.adsabs.harvard.edu/abs/2007A&A...461..823V} {461, 823}

\bibitem[\protect\citeauthoryear{{Wang} et~al.,}{{Wang}
  et~al.}{2016}]{Wang2016}
{Wang} T.,  et~al., 2016, \mn@doi [\apj] {10.3847/0004-637X/828/1/56}, \href
  {https://ui.adsabs.harvard.edu/abs/2016ApJ...828...56W} {828, 56}

\bibitem[\protect\citeauthoryear{{Weinberger} et~al.,}{{Weinberger}
  et~al.}{2017}]{Weinberger2017}
{Weinberger} R.,  et~al., 2017, \mn@doi [\mnras] {10.1093/mnras/stw2944}, \href
  {https://ui.adsabs.harvard.edu/abs/2017MNRAS.465.3291W} {465, 3291}

\bibitem[\protect\citeauthoryear{{Williams}, {Quadri}, {Franx}, {van Dokkum}
  \& {Labb{\'e}}}{{Williams} et~al.}{2009}]{Williams2009}
{Williams} R.~J.,  {Quadri} R.~F.,  {Franx} M.,  {van Dokkum} P.,   {Labb{\'e}}
  I.,  2009, \mn@doi [\apj] {10.1088/0004-637X/691/2/1879}, \href
  {https://ui.adsabs.harvard.edu/abs/2009ApJ...691.1879W} {691, 1879}

\bibitem[\protect\citeauthoryear{{Willis} et~al.,}{{Willis}
  et~al.}{2020}]{Willis2020}
{Willis} J.~P.,  et~al., 2020, \mn@doi [\nat] {10.1038/s41586-019-1829-4},
  \href {https://ui.adsabs.harvard.edu/abs/2020Natur.577...39W} {577, 39}

\bibitem[\protect\citeauthoryear{{Wylezalek} et~al.,}{{Wylezalek}
  et~al.}{2013}]{Wylezalek2013}
{Wylezalek} D.,  et~al., 2013, \mn@doi [\apj] {10.1088/0004-637X/769/1/79},
  \href {https://ui.adsabs.harvard.edu/abs/2013ApJ...769...79W} {769, 79}

\bibitem[\protect\citeauthoryear{{Wylezalek} et~al.,}{{Wylezalek}
  et~al.}{2014}]{Wylezalek2014}
{Wylezalek} D.,  et~al., 2014, \mn@doi [\apj] {10.1088/0004-637X/786/1/17},
  \href {https://ui.adsabs.harvard.edu/abs/2014ApJ...786...17W} {786, 17}

\bibitem[\protect\citeauthoryear{{van der Burg} et~al.,}{{van der Burg}
  et~al.}{2013}]{VanderBurg2013}
{van der Burg} R.~F.~J.,  et~al., 2013, \mn@doi [\aap]
  {10.1051/0004-6361/201321237}, \href
  {https://ui.adsabs.harvard.edu/abs/2013A&A...557A..15V} {557, A15}

\bibitem[\protect\citeauthoryear{{van der Burg}, {McGee}, {Aussel}, {Dahle},
  {Arnaud}, {Pratt}  \& {Muzzin}}{{van der Burg} et~al.}{2018}]{VanderBurg2018}
{van der Burg} R. F.~J.,  {McGee} S.,  {Aussel} H.,  {Dahle} H.,  {Arnaud} M.,
  {Pratt} G.~W.,   {Muzzin} A.,  2018, \mn@doi [\aap]
  {10.1051/0004-6361/201833572}, \href
  {https://ui.adsabs.harvard.edu/abs/2018A&A...618A.140V} {618, A140}

\bibitem[\protect\citeauthoryear{{van der Walt}, {Colbert}  \&
  {Varoquaux}}{{van der Walt} et~al.}{2011}]{numpy:2011}
{van der Walt} S.,  {Colbert} S.~C.,   {Varoquaux} G.,  2011, \mn@doi
  [Computing in Science \& Engineering] {10.1109/MCSE.2011.37}, 13, 22

\makeatother
\end{thebibliography}



\appendix


\bsp	
\label{lastpage}
\end{document}